\documentclass[11pt]{article}

\usepackage{epsfig}
\usepackage{epstopdf}
\usepackage{color}
\usepackage{setspace}
\usepackage[margin=0.8in]{geometry}
\usepackage{latexsym}
\usepackage{authblk}
\usepackage{amsmath}
\usepackage{amsfonts}
\usepackage{amssymb}
\usepackage{psfrag}
\usepackage{graphicx}
\usepackage{algorithm2e}

\usepackage{lineno}
\title{Inverse Problem for Dynamic Computer Simulators via Multiple Scalar-valued Contour Estimation}

\author{Joseph Resch\\University of California, Los Angeles, USA            
	 \and Abhyuday Mandal\\University of Georgia, Athens, USA \and Pritam Ranjan\\OM\&QT, Indian Institute of Management Indore, MP, India}

\date{}

\begin{document}

	\maketitle

	{\doublespacing
		\begin{abstract}
			
			\noindent In this paper we consider a dynamic computer simulator that produces a time-series response $y_t(x)$ over $L$ time points, for every given input parameter $x$. We propose a method for solving inverse problems, which refer to the finding of a set of inputs that generates a pre-specified simulator output. Inspired by the sequential approach of contour estimation via expected improvement criterion developed by Ranjan et al. (2008,  DOI: 10.1198/004017008000000541), our proposed method discretizes the target response series on $k \; (\ll L)$ time points, and then iteratively solves $k$ scalar-valued inverse problems with respect to the discretized targets.  We also propose to use spline smoothing of the target response series to identify the optimal number of knots, $k$, and the actual location of the knots for discretization. The performance of the proposed methods is compared for several test-function based computer simulators and the motivating real application that uses a rainfall-runoff measurement model named Matlab-Simulink model.
		\end{abstract}
	}


	%	\clearpage

	%%%%%%%%%%%%%%%%%%%%%%%%%%%%%%%%%%%%
	%                                                                                             %
	%                                                                                             %
	%                                                                                             %
	%                                                                                             %
	%                                                                                             %
	%                             SECTION 1: Introduction                          %
	%                                                                                             %
	%                                                                                             %
	%                                                                                             %
	%                                                                                             %
	%                                                                                             %
	%%%%%%%%%%%%%%%%%%%%%%%%%%%%%%%%%%%%

	\clearpage \doublespacing
	
\section{INTRODUCTION}

%1. context 2. problem statement 3. why it is important 4. what has been done 5. What are we proposing 6. What's novel here. 

Physical experiments are frequently expensive and impractical to perform. The growth in computing power during modern times offers an alternative to carry out such experiments in the real world scenario. Due to the high cost of resources that come with conducting physical experiments, less expensive computer simulators are used to represent phenomena, such as, dynamic traffic patterns of a metropolitan intersection, hydrological behaviors of an ecosystem, the spread behaviour of a wildfire, formation patterns of galaxies, and so on. The applications of computer simulation models span a variety of sectors including ecology, medicine, engineering, industrial experiments, nuclear research, manufacturing, climatology, and astronomy. {Complex computer experiments, although much less expensive than physical experiments, are often still computationally expensive, and cost-efficient surrogate simulators are called for.}

In this paper we focus on the calibration of expensive to evaluate dynamic computer simulators. That is, the simulator outputs are time-series and the objective is to solve the inverse problem (also sometimes called as the calibration problem) which attempts to find the input parameters of the simulator that generate either exact or close approximation of a pre-specified target. The application that motivated this study comes from a rainfall-runoff measurement model called Matlab-Simulink model which predicts the rate of runoff and sediment yield (Duncan et al., 2013). Here, the objective is to find the input parameters of the Matlab-Simulink model that generate simulator outputs as close as possible to the real data collected from a watershed from the Bioconversion center at the University of Georgia, Athens, USA.

Since computer experiments tend to be complex and have high dimensional inputs, direct attempt to solve the inverse problem using simulator outputs alone is infeasible. Thus, less computationally expensive surrogates such as the Gaussian Process models are fitted using training data of $n$ observations $(x_1, y_1), \dots, (x_n, y_n)$ to emulate computer models. During the past few years, the inverse problem for expensive to evaluate complex computer models with \emph{scalar-valued responses} has been given extensive focus (e.g. Bingham et al., 2014; Picheny et al., 2013; Ranjan et al., 2008). In contrast, approaches to solve the calibration problem for \emph{dynamic computer models} have been less studied. Vernon et al. (2010) selected a handful of time-points from the target response series (called discretization point set (DPS)) and then developed a batch-sequential approach called the history matching (HM) to simultaneously solve multiple scalar-valued inverse problems which would approximate the underlying dynamic inverse problem, but the method required too many simulator runs. Recently, Bhattacharjee et al. (2019) proposed a modification which required fewer simulator runs for the calibration of hydrological simulators.  Ranjan et al. (2016) suggested a scalarization approach to efficiently minimize $\|g(x)-g_0\|$ via the expected improvement based sequential strategy developed by Jones et al. (1998), where $g(x)=\{g(x,t_j), j=1,2,...,L\}$ is the simulator output for the input $x$ and $g_0=\{g_0(t_j), j=1,2,...,L\}$ is the target response. Zhang et al. (2019) used a singular value decomposition based Gaussian process model to fit a surrogate for the dynamic simulator and then generalized the expected improvement approach of Jones et al. (1998) for minimizing $\|g(x)-g_0\|$.

Both Vernon et al. (2010) and Bhattacharjee et al. (2019) used an adhoc method (or a subjective expert opinion) for choosing DPS, and then used an implausibility criterion (similar in spirit as the improvement function) to sort through the input space and find the inverse solution. In this paper, we propose using cubic spline smoothing of the target series to systematically construct the DPS as the optimal knots identified in a sequential manner inspired by the forward variable selection method. This method allows us to identify both the size and positions of the DPS and divide the predetermined budget accordingly.  Subsequently, we recommend solving the scalar-valued inverse problems iteratively using the contour estimation method developed by Ranjan et al. (2008). Finally, the inverse solution of the underlying dynamic simulator is obtained by taking the intersection of the solutions from the scalar-valued inverse problems.

The remaining sections are outlined as follows. Section 2 reviews the concepts integral to the proposed method and review the competing modified history matching (HM) approach proposed by Bhattacharjee et al. (2019). Section 3 provides the elements of the proposed multiple scalar contour estimation (MSCE) method along with thorough implementation details of the key steps. In Section 4, we establish the superiority of the proposed method via three test functions based simulator. Section~5 discusses the real motivating Matlab-Simulink application. We provide some concluding remarks in Section 6.

\section{REVIEW OF EXISTING METHODOLOGY}
In this section, the existing methods that set precedence for this paper are reviewed. We examine the use of a GP model as a surrogate to a deterministic simulator and the use of the expected improvement criterion for choosing the follow-up trials in the sequential design framework. We also review the modified history matching approach of Bhattacharjee et al. (2019) and the use of a discretization-point-set (DPS) to approximate the inverse solution for dynamic computer simulators. 

%\textcolor{red}{To fix notations, please define $x$, $\chi$, $s$ etc. for the rest of the paper here. We have $s$ on equation (4), on equation (6) and on equation (7). Are these same $s$'s?} \textcolor{blue}{These are all either fixed or elaborated on.}

\subsection{Gaussian Process Models}
Deterministic computer model simulators of complex phenomena are often computationally expensive to evaluate, and hence the emulation via a statistical surrogate becomes much more practical. A useful surrogate for this purpose is the Gaussian Process (GP) model (Sacks et al., 1989). For a set of input-output combinations, a stationary GP model, called the ordinary Kriging, assumes:
\begin{eqnarray*}
	y(x_i) = \mu + Z(x_i), \hspace{0.5in} i = 1, \dots, n,
\end{eqnarray*}
where $\mu$ is the mean and $Z(x_i)$ is a Gaussian Process with $E(Z(x_i)) = 0$ and a covariance structure of $Cov(Z(x_i), Z(x_j)) = \sigma^2R(\theta; x_i, x_j)$. There are several popular choices of $R(\cdot,\cdot)$. The power-exponential correlation structure will have the $(i,j)^{\mbox{th}}$ term $R_{ij}(\theta)$ as:
\begin{eqnarray}\label{eq:1}
R(Z(x_i), Z(x_j)) = \prod_{k = 1}^{d} \exp\Bigg\{-\theta_k \mid x_{ik} - x_{jk} \mid ^{p_k} \Bigg\} \hspace{0.2in} \mbox{for all }  i,j,
\end{eqnarray}
where $0 < p_k \leq 2$ are smoothness parameters and $\theta_k$ measure correlation strength. In this paper, we assume power exponential correlation with $p_k = 1.95$ (for numerical stability and smoothness). The best linear unbiased predictor for the response at any unsampled point $x^*$ is given by:
\begin{eqnarray*}
	\hat{y}(x^*) = \hat{\mu} + r(x^*)^T R_n^{-1}(y-1_n \hat{\mu}),
\end{eqnarray*}
where $r(x^*) = \Big[ \mbox{corr} (z(x^*), z(x_1)), \dots, \mbox{corr} (z(x^*), z(x_n))\Big]^{\tt T}$, $R_n$ is the $n\times n$ correlation matrix with elements $R_{ij}$ (as seen in Equation (\ref{eq:1})), and the prediction uncertainty is quantified by
\begin{eqnarray}\label{eq:2}
s^2(x^*) = \hat{\sigma}^2 \bigg(1-r(x^*)^T R_n^{-1}r(x^*) \bigg).
\end{eqnarray}

The flexibility of the correlation structure makes the GP model a popular surrogate for complex computer models. Throughout this paper, the \textit{R} package {\tt GPfit} (MacDonald et al., 2015) is used to fit GP models.

\subsection{Sequential Design}
For finding optimal solutions of an inverse problem for computationally intensive computer simulators, sequential designs have been proven to outperform the competing popular approaches, e.g., Ranjan et al. (2008), Ranjan et al. (2016), and Zhang et al. (2019). The basic framework remains the same as in the global optimization developed by Jones et al. (1998) - which consists of two components, finding a good initial design and then sequentially choose the follow-up trial locations for simulation runs.

In computer experiments, space-filling designs are popular choices for the initial design. Suppose we have $d$ input variables, a space-filling design such as a maximum projection Latin hypercube design (Joseph et al., 2015), is used to create a training input set of initial size $n_{0}$ from the scaled input space $[0,1]^{d}$. The corresponding responses are generated by evaluating the simulator at each input of the training set. A surrogate model is then fitted to the simulator responses and the inputs. After which, a sequential design criterion such as expected improvement (EI) is evaluated using the GP model over the entire input space to find the input, $x_{new}$, that leads to the greatest expected improvement (see Jones et al. (1998) and Bingham et al. (2014) for details). Practically, we evaluate EI over a dense, randomly generated spacing-filling test set of large size $M$ in $[0,1]^{d}$. The $x_{new}$ and corresponding true simulator response are augmented to the training set. The surrogate (GP model) is refitted to a new training set. This iterative process, of optimizing EI to choose $x_{new}$ and refitting the surrogate to the augmented data, is repeated until the total budget of $N$ points is exhausted. The optimal inverse solution would be extracted from the final fit.

\subsection{Expected Improvement Criterion for Contour Estimation}

Jones et al. (1998) proposed the first EI criterion for global optimization for scalar-valued simulators.  In the same spirit, Ranjan et al. (2008) developed a EI criterion for estimating the inputs that lead to a pre-specified scalar target response. The corresponding improvement function is given by 
\begin{equation*}\label{Eq.EI_contour1}
I(x^{*}) = \epsilon^{2}(x^{*}) - \mbox{min}\Big[ \{ y(x^{*}) - a\}^{2}, \epsilon^{2}(x^{*}) \Big].
\end{equation*}
where $\epsilon(x^{*}) = \alpha s(x^{*})$ for a positive constant $\alpha$ (e.g., $\alpha = 0.67$, corresponding to 50\% confidence under approximate normality), $s(x^*)$ is defined in (\ref{eq:2}), and $a$ is the pre-specified target response.  Hence, the EI value (which is simply the expected value of the improvement function) is:
\begin{eqnarray}\label{Eq.EI_contour2}
E[I(x^{*})] =& \Big[\epsilon^{2}(x^{*}) - \{ \hat{y}(x^{*}) - a \}^{2} \Big] \Big\{\Phi(u_{2}) - \Phi(u_{1})\Big\} \nonumber \\
&+ s^{2}(x^{*})\Big[ \{u_{2} \phi(u_{2}) - u_{1} \phi(u_{1})\} \{\Phi(u_{2}) - \Phi(u_{1})\}\Big] \nonumber \\
&+ 2\Big\{ \hat{y}(x^{*}) - a \Big\}s(x^{*}) \Big\{\phi(u_{2}) - \phi(u_{1})\Big\},
\end{eqnarray}
where $u_{1} = [a - \hat{y}(x^{*}) - \epsilon(x^{*})] / s(x^{*})$, and $u_{2} = [a - \hat{y}(x^{*}) + \epsilon(x^{*})] / s(x^{*})$, and $\Phi$, $\phi$ are the cumulative distribution function and probability density function of the standard normal distribution for input $x^{*}$.   

The EI based selection strategy for follow-up trials became extremely popular because different components of EI focus on two separate aspects, (1) exploration of the input space which prohibits getting stuck at a local optimum, and (2) exploiting the areas of interest for more precise estimate of the objective.

\subsection{Modified History Matching for the Inverse Problem} \label{hm_via_sd}

Recall that $g(x) = \{g(x, t_j), j = 1, \dots, L\}$ denote the time-series valued simulator response with the input scaled in $[0,1]^d$. The aim of the inverse problem (or calibration problem) is to find $x$ such that $g(x)$ resembles the target response $g_0 = \{g_0(t_j), j = 1, \dots, L\}$. To this end, a history matching approach was first proposed by Vernon et al. (2010) and was subsequently modified by Bhattacharjee et al. (2019) to solve the inverse problem for dynamic simulators. One of the key modifications in Bhattacharjee et al. (2019) was to use a small space-filling Maximin Latin hypercube design (Johnson et al., 1990) as compared to a large Latin hypercube design for fitting the initial GP surrogate.

Although the simulator output is a time-series of length $L$,  the history matching approach selects a handful of time-points, which are referred to as a discretization-point-set (DPS) and has size $k\ll L$.  The said approach uses the simulator output at only the DPS time-points ($t_1^*, t_2^*, \dots, t_{k}^*)$ to solve the underlying inverse problem. That is, the reduced objective is to finds $x$ such that $g(x, t_j^*) = g_0(t_j^*)$ for all $j = 1, \dots, k$. As a result, it is important for the DPS to be representative of the simulator response. 

The history matching uses a multi-stage design approach to augment the training set. At each stage, a criterion called the implausibility function is evaluated over a large test set from the input space ($\chi$), and then the design points are said to be implausible if the criterion value exceeds certain pre-determined cutoff. Subsequently, the plausible points, $\{ x \in \chi : IM_{max}(x) \leq c\}$, are augmented to the training set. { For each $j = 1, \dots, k$, the}  implausibility criterion is defined as
\begin{equation*}
IM_{j}(x) = \frac{\mid\hat{g}(x, t_j^*) - g_0(t_j^*)\mid}{s_{t_j^*}(x)},
\end{equation*}
where $\hat{g}(x, t_j^*)$ is the predicted response derived from the GP surrogate corresponding to the simulator response at time point $t_j^*$ and $s_{t_j^*}(x)$ is the associated uncertainty. Design points are deemed implausible if $IM_{max}(x) > c$, where $c$ is the pre-determined cutoff chosen in an ad hoc manner and
\begin{eqnarray*}
	IM_{max}(x) = \mbox{max}\{IM_{1}(x), IM_{2}(x), \dots, IM_{k}(x)\}.
\end{eqnarray*}

Following each stage of training data augmentation, the GP surrogate is updated. This iterative process of updating the surrogate fit and selecting the plausible points,   continues until no additional points are to be added from the test set. At the end of the procedure, the best approximate inverse solution is extracted from the training set or from the final GP surrogate.

\section{Proposed Methodology}

We propose to solve this dynamic computer simulator inverse problem via multiple scalar-valued inverse problems under a limited budget constraint. Similar to the (modified) history matching method, we discretize the simulator response at a DPS of size $k (\ll L)$ that aims to capture the important features of the time-series response.  Subsequently, we solve the $k$ scalar-valued inverse problems using the contour estimation method developed by Ranjan et al. (2008) discussed in Sections~2.2 and 2.3. Finally, the intersection of these $k$ sets of inverse solutions is taken as the optimal solution for the original inverse problem for dynamic computer simulator. 

There are several parts of the proposed methodology that requires detailed discussion. First, we present the construction of the DPS. Both the history matching approach by Vernon et al. (2010) and the modified history matching technique by Bhattacharjee et al. (2019) choose DPS rather subjectively, and do not follow any systematic algorithm. Here we propose an algorithmic approach for choosing the DPS.

We propose fitting cubic splines to the target response series and then use the set of knot locations as the DPS. Undoubtedly, finding optimal number and location of knots is a challenging problem in spline regression. We suggest an iterative but greedy approach for constructing the DPS. The idea is similar to the construction of a regression tree, where the split-points are essentially the knot locations. That is, we start with no knots, and find the best location for the first knot by minimizing the overall mean squared error as per the spline regression fit. The optimal location for the second knot is found by fixing the first knot location. Continuing further in this manner, the search for optimal location for the $j$-th knot assumes that the optimal location of the previous $j-1$ knots are known. Finally, the optimal number of knots are found using something similar to scree-plot, where we plot  mean squared error against the number of knots and identify the elbow of the plot. For implementation, the R package {\tt splines} is called upon for this purpose while the command {\tt bs()} is used for finding B-spline basis functions in the linear model environment.

We quickly illustrate the details by applying it to a test function. Suppose the simulator outputs are generated via Easom function (Michalewicz, 1996), defined as,
\begin{equation*}
g(x,t_j) = \cos(x_1)\cos(x_2) \exp\big\{-(x_1-\pi t_j)^2-(x_2-\pi)^2)\big\},
\end{equation*}
where $t_j$ are $L$ equidistant time points in $[0, 1]$ for $j = 1, \dots, L = 200$, and the input space is scaled to $(x_1, x_2) \in [0, 1]^2$. We select the target response $g_0$ corresponding to the input set $x_0 = (0.8, 0.2)$. Here the objective is to solve the inverse problem by calibrating $x=(x_1,x_2)$ such that $g(x,t_j) \approx g_0(t_j)$ for all $j$.

The first element of the DPS is obtained by minimizing the mean square errors of the cubic spline fits onto the target response over each of the possible 200 time points as the sole knot. We found the optimal first knot at time point $t_1^*=145$. Keeping the knot at time point $t_1^*=145$ fixed, we repeated the process and found the second knot at time point $t_2^*=37$. The process continued, and the locations of ten optimal knot are $\{145, 37, 132, 47, 120, 55, 113, 63, 104, 174\}$ (see Figure~1). %We can see the goodness-of-fit using the first one through ten knots in this list to fit cubic splines onto the target response in Figure 1.

\begin{figure}[h!]\centering
	\includegraphics[scale=0.75]{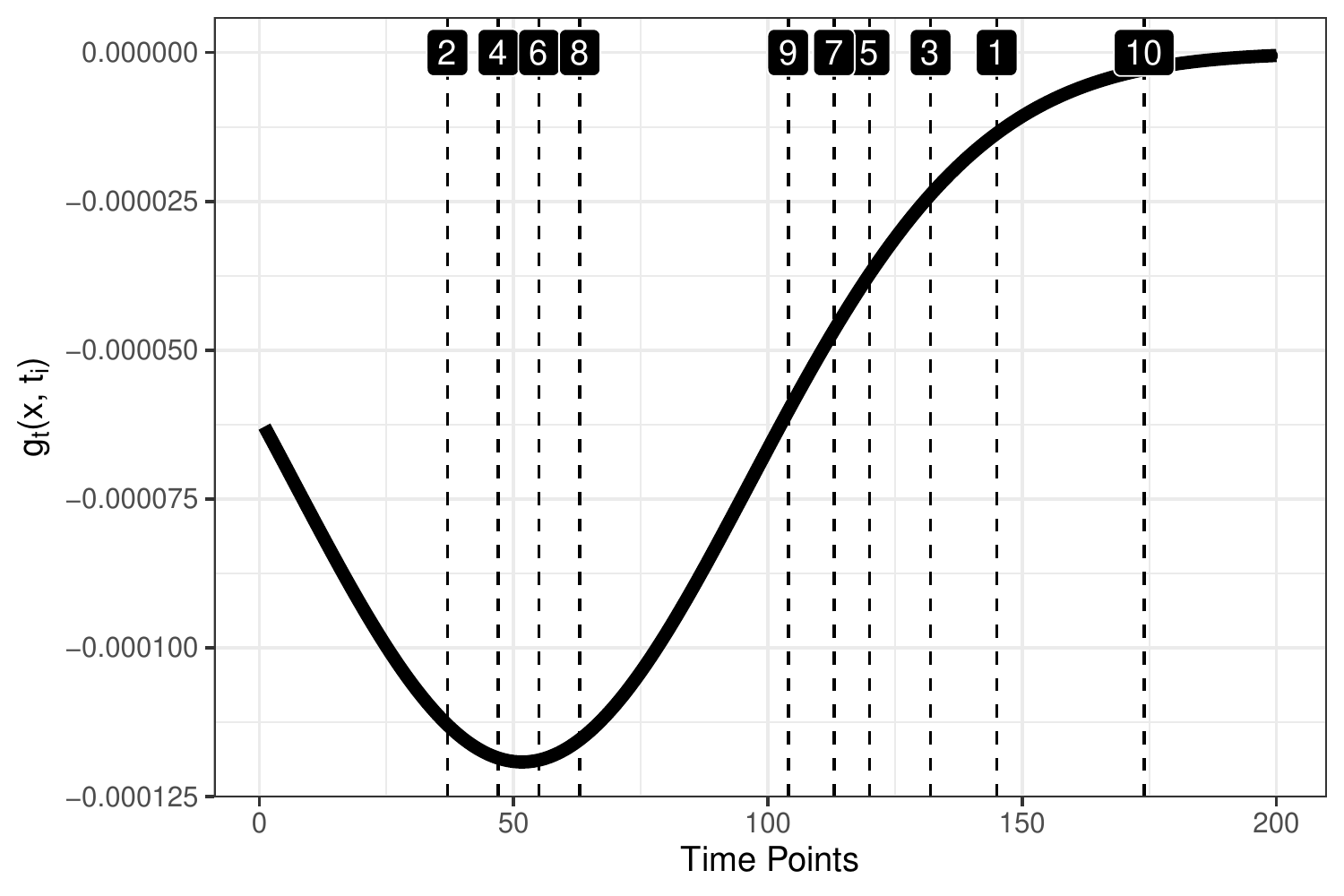} 
	\caption{Easom function: The vertical dashed lines depict the ordered positioning of optimal knots for fitting cubic splines to the time-series.}
\end{figure}

In Figure~1, we have added 10 knots, however, in reality, the required number of knots may be different. Following the idea of scree-plot from the principal component analysis, we look at the mean squared error (MSE) versus the number of knots function, and try to find an ``elbow" in the plot to find the optimal number of knots. The idea is to find the point in the ``MSE vs. the number of knots function" where the second derivative reaches a positive value. This would allow for a good fit while maintaining the efficiency of the knots used. Figure~2 shows the corresponding ``MSE vs. the number of knots function" plot for the Easom function. In this case, the elbow cutoff is $3$. That is, the required discretization-point-set (DPS) for this time series response would be $\{145, 37, 132\}$.

\begin{figure}[h!]\centering
	\includegraphics[scale=0.75]{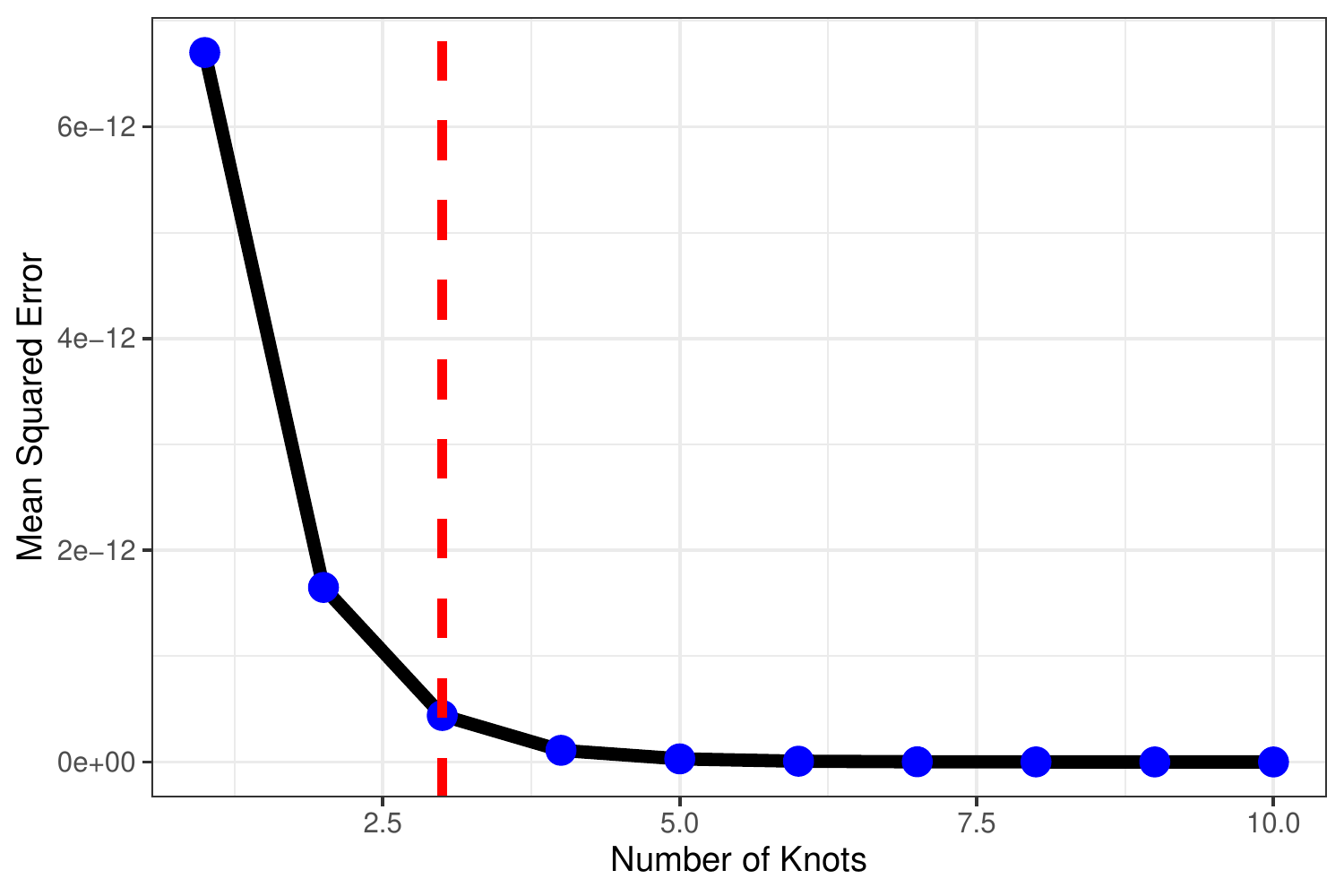} 
	\caption{Easom function: ``Mean squared error versus the number of knots" for 10 knots for spline regression added sequentially one at-a-time. }
\end{figure}

After finding a reasonable DPS, we sequentially solve $k$ scalar-valued inverse problem via contour estimation approach developed by Ranjan et al. (2008). Suppose our budget for the simulator runs is $N$, then, the process starts by  choosing an initial design of size $n_0 (<N)$ from the input space $[0,1]^d$.  We use a space-filling, maximum projection Latin hypercube design (Joseph et al., 2015), for the initial design. The remainder of the budget $N-n_0$ is equally distributed for estimating the $k$ scalar-valued inverse solutions. That is, the first inverse problem would estimate $S_1(x) = \{x: g(x,t_1^*)=g_0(t_1^*)\}$ using $n_0$-point initial design and $(N-n_0)/k$ follow-up trials chosen sequentially by maximizing the EI criterion based on the GP surrogate given by equation~(\ref{Eq.EI_contour2}) in Section~2.3. The augmented data are now treated as the initial training set for the second scalar-valued inverse problem. Thus, one would estimate $S_j(x) = \{x: g(x,t_j^*)=g_0(t_j^*)\}$ using the initial training set of size $n_0 + (j-1)(N-n_0)/k$, obtained after solving the previous $j-1$ scalar-valued inverse problems, and $(N-n_0)/k$ sequential trials via EI optimization.

For the Easom function, since the DPS is of size three, we need to solve three scalar-valued inverse problems. We set a total training size budget of $N=50$ points and initial design of size $n_0 = 15$. The budget of follow-up points, $N - n_0 = 35$, is divided approximately evenly for the three inverse problems. When computing the EI criterion, we set $\alpha = 0.67$ which corresponds to $50\%$ confidence interval under normality. Furthermore, since the input space is only two-dimensional unit square, we use $5000$-point random Latin hypercube designs for maximizing the EI criteria for sequentially adding follow-up trials. Figure~3 shows the three estimated contours along with selected follow-up points corresponding to $t_1^*=145$ (in red), $t_2^*=37$ (in green) and $t_3^* = 132$ (in blue).

\begin{figure}[h!]\centering
	\includegraphics[scale=0.7]{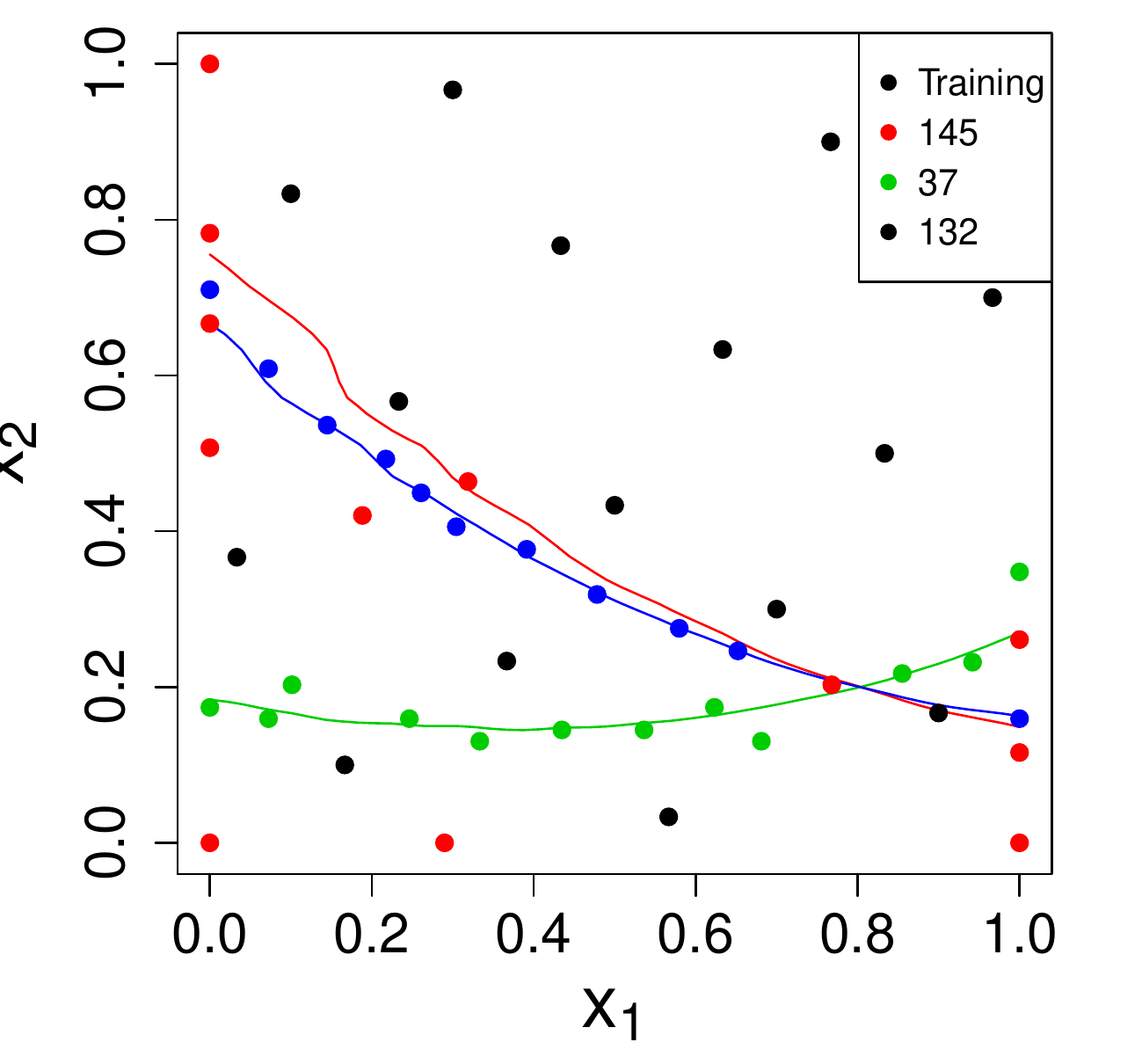}  
	\caption{Easom function: Training data is depicted by dots and the estimated contours are shown by solid curves. The black dots correspond to the initial design, whereas red, green, and blue dots represent the follow-up points obtained via EI optimization for the three scalar-valued contour estimation at $DPS =(145,37,132)$, respectively.}
\end{figure}

From Figure~3, it is clear that for the first contour estimation, more follow-up points focus on global exploration for better overall understanding of the process as compared to local search for accuracy enhancement of the contour estimate.

The intersection of $k$ scalar-valued inverse problem solutions at the time points of the discretization-point-set (DPS) is used as the inverse solution of the underlying dynamic simulator. For Easom function example, it is clear from Figure~3, that the three contours intersect on a common point which is the true inverse solution $x_0=(0.8, 0.2)$.  To implement this, final GP model surrogate fitted after the final iteration is used to extract $x_{opt}$. Instead of the exact match, we accept the inverse solutions as  $S_i = \{x^{*}: |\hat{g}(x^*, t_{i}^*) - g_{0}(t_i^*)| < \epsilon\}$, for each time point $t_i^*$ in DPS. This accounts for the round off errors and other approximations made during the implementation. This tolerance $\epsilon$ has to be subjectively decided to accurately estimate the inverse solution set. Assuming that the solution exists and $\cap_{i=1}^kS_i$ represents a single contiguous region, one can find $x_{opt} = argmin\{\|g(x)-g_0\|, x \in \cap_{i=1}^kS_i\}$. For Easom function example, we set $\epsilon = 10^{-5}$, and the final inverse solution obtained is $x_{opt} = (0.8188, 0.2029)$. Figure 4 shows that the simulator response at $x_{opt}$ is virtually indistinguishable as compared to the target response. 

%\textcolor{red}{(What happens if $\cap_{i=1}^kS_i$ is not a single contiguous region? Need to discuss that possibility.)} \textcolor{blue}{Note: If the contiguous region does not exist, a solution does not for the dynamic simulator.}

\begin{figure}[h!]\centering
	\includegraphics[scale=0.8]{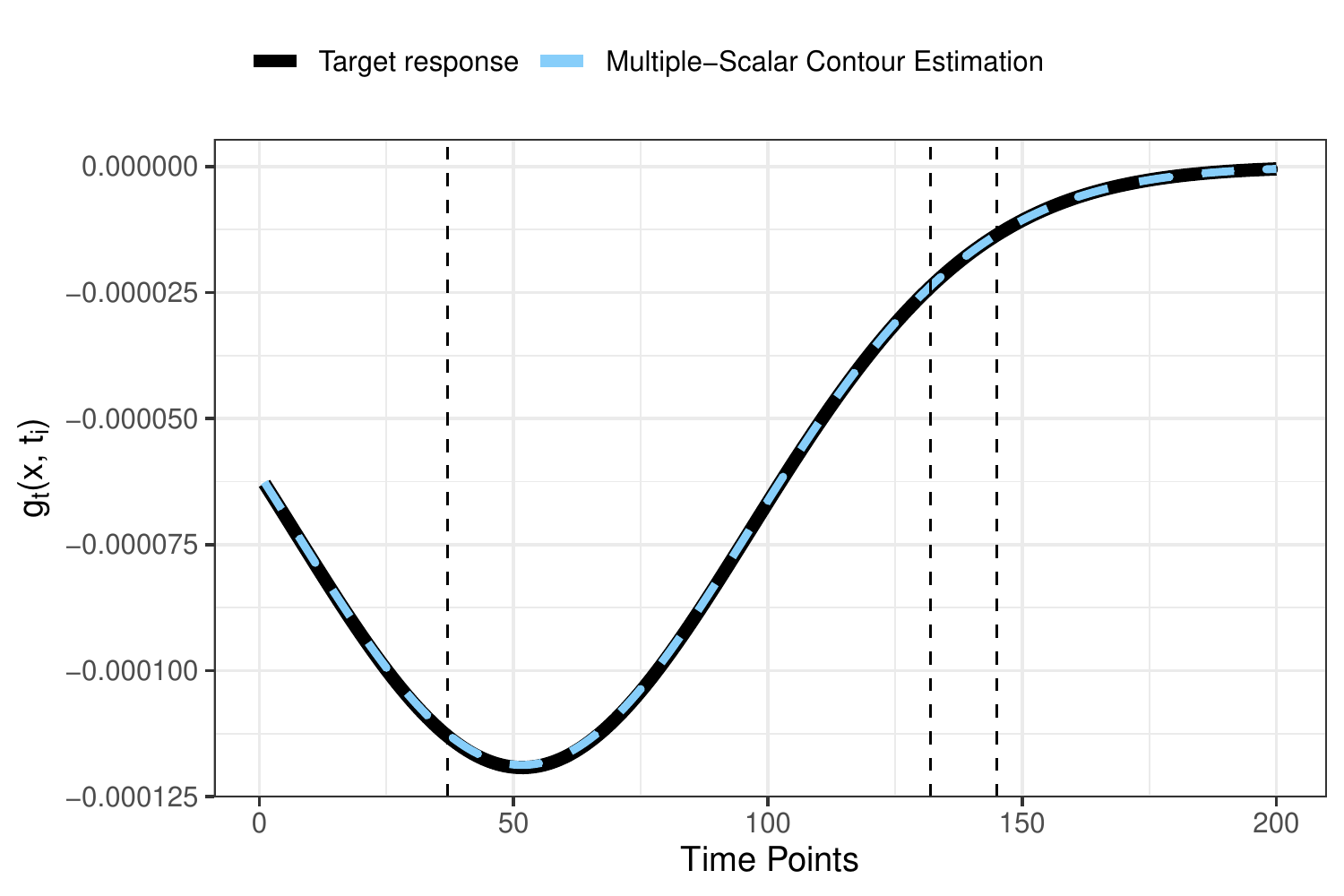}  
	\caption{Easom function: The target response is shown by the solid black line and the simulator response at $x_{opt}$ is shown by the dashed blue curve.}
\end{figure}

\noindent We now summarize the key steps of the proposed approach in Algorithm~1.

{\small
	\begin{algorithm}[t!]
		%\setstretch{1.4}
		\SetKwInOut{Input}{Input}\SetKwInOut{Output}{Output}
		\DontPrintSemicolon
		%\SetAlgoLined
		\Input{(1) Input parameters: $n_0, d, L, N$ \\ (2) Target response: $\{g_0(t_j),j=1,...,L\}$\\ (3) Tolerance: $\epsilon$ \\ (4) Dynamic computer simulator: $\{g(x,t_j), j=1,...,L\}$}
		\Output{(1) Final training set: $\texttt{xx}_{N\times d}$ and $\texttt{yy}_{L\times N}$\\ (2) Estimated inverse solution: $\texttt{x}_{opt}$}
		%\BlankLine
		\hrule
		Construct a DPS of size $k (\ll L)$ that would capture the important features of the target time-series response, say, $(t_1^*, t_2^*, ..., t_k^*)$.	See Section~3 for the proposed spline based methodology. \;
		Choose $n_0$ points in $[0,1]^d$ using maximin Latin hypercube design. Obtain the corresponding simulator response matrix $Y_{L\times n_0}$.\;
		\For{$j = 1, \ldots, {k}$}{
			Use scalar-valued contour estimation method by Ranjan et al. (2008) to estimate $S_j(x)=\{x\in \chi \ : \ |g(x,t_j^*) - g_0(t_j^*)| < \epsilon \}$. Assume the size of initial design is $n_0 + (j-1)\cdot (N-n_0)/k$, whereas $(N-n_0)/k$ follow-up trials are added sequentially one at-a-time as per the EI criterion in Section~2.3. \;
			Augment the follow-up points to the initial design for the next scalar-valued inverse problem.\;
		}
		Extract the final inverse solution as $\cap_{j=1}^{k} S_i(x)$.\;
		\caption{Multiple scalar-valued contour estimation approach}
	\end{algorithm}
}

%\clearpage
\section{Simulation studies}

In this section, we use three different test simulators to compare the performance of the proposed method with the modified history matching algorithm. Since the number of simulator runs cannot be fixed beforehand in the latter approach, we implement the proposed multiple scalar-valued contour estimation method at two settings - using a prefixed limited budget, and the budget matching with that of the competing method. For performance comparison between the two methods, we use three popular goodness of fit measures called $R^2$, RMSE and normD. The objective would be to maximize $R^2$ and minimize RMSE, normD.  

%\textcolor{red}{(Please mention what $L$ is in the context of the four measures below.)} \textcolor{blue}{The notation for the dynamic simulator $\{g(x,t), t=1,...,L\}$ is reintroduced in the equations.}

\begin{itemize}
	
	\item Root mean squared error
	\begin{eqnarray*}
		RMSE = \left(\frac{1}{L} \sum_{j=1}^{L} \mid g(\hat{x}_{opt}, t_j) - g_0(t_j) \mid ^2 \right)^{1/2}.
	\end{eqnarray*}
	
	\item Coefficient of determination $R^2$ of the simple linear regression model fitted to the estimated inverse solution and the target response, i.e., $R^2$ of the following linear regression model:
	\begin{eqnarray*}
		g_0(t_j) = g(\hat{x}_{opt}, t_j) + \delta_j, j = 1, 2, \dots, L, 
	\end{eqnarray*}
	with the assumption of i.i.d. errors $\delta_j$.

	\item Normalized discrepancy (on log-scale), between the simulator response at the estimated inverse solution and the target response
	\begin{eqnarray*}
		normD=\log\left(\frac{\left\|{g_0}-{g}\left(\hat{{x}}^{*}\right)\right\|_{2}^{2}} {\left\|{g_0}-\bar{g}_0 1_{L}\right\|_{2}^{2}}\right)
	\end{eqnarray*}
	where $\bar{g}_0 = \sum_{t=1}^{L}g_0(t)/L$ and $1_{L}$ is an L-dimension vector of ones. Note that $1-\exp(normD)$, also referred to as Nash–Sutcliffe Efficiency (Nash and Sutcliffe, 1970), is a popular goodness of fit measure in hydrology literature.
\end{itemize}

\subsection{Example 1: Easom Function (Michalewicz, 1996) continued}
We begin by revisiting the example discussed in Section 3 to compare the inverse solutions arrived at by the two methods. The modified history matching approach used a budget size of $N = 230$ when using a predetermined cutoff of $c = 0.5$ (for implausibility measure). Thus, the proposed multiple scalar-valued contour estimation (MSCE) was implemented in two scenarios: a budget of the matching size, and the original budget size of $N = 50$.

\begin{table}[h!]
	\centering
	\caption{Easom Function: Performance comparison of the proposed MSCE method with $N=50$ and $N=230$, and the modified history matching (HM) method which required $N=230$ simulator runs. } 
	\medskip
	\begin{tabular}{ccccc}
		\hline
		\bf{Methods} & \bf{Total Budget} & \bf{RMSE} & \bf{R$^2$}  & \bf{normD}\\
		\hline
		HM & $N = 230$ & $2.46\times10^{-7}$ & 0.999 &  $3.157\times10^{-5}$\\
		%\hline
		MSCE & $N = 50$ & $3.10\times10^{-7}$ & 0.999 &  $4.993\times10^{-5}$\\
		%\hline
		MSCE & $N = 230$ & $1.48\times10^{-7}$ & 0.999 &  $1.137\times10^{-5}$\\
		\hline
	\end{tabular}
	\label{Tab: easom1}
\end{table}

\begin{figure}[h!]\centering
	\begin{tabular}{lll}
		\includegraphics[scale=0.3]{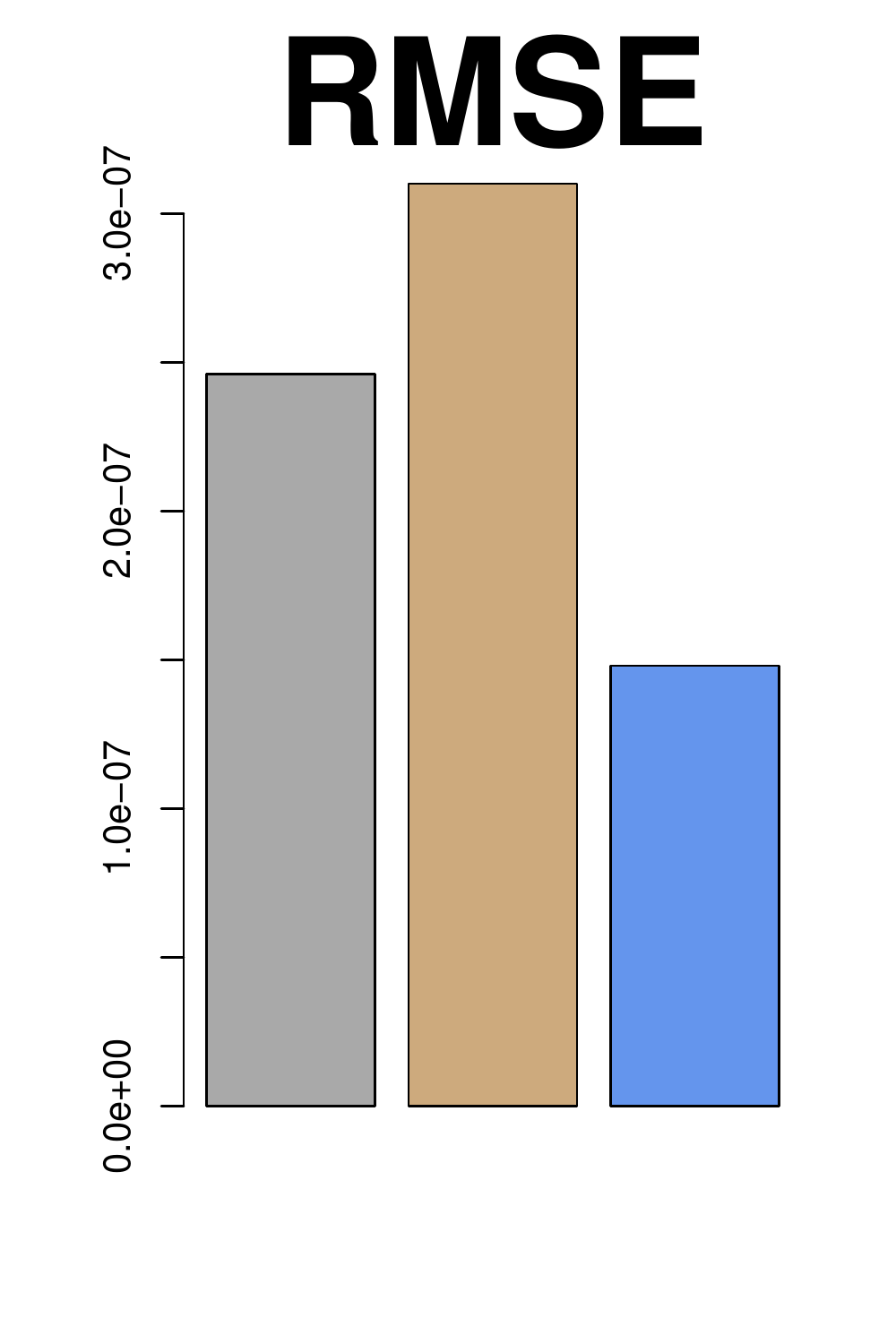} &
		\includegraphics[scale=0.3]{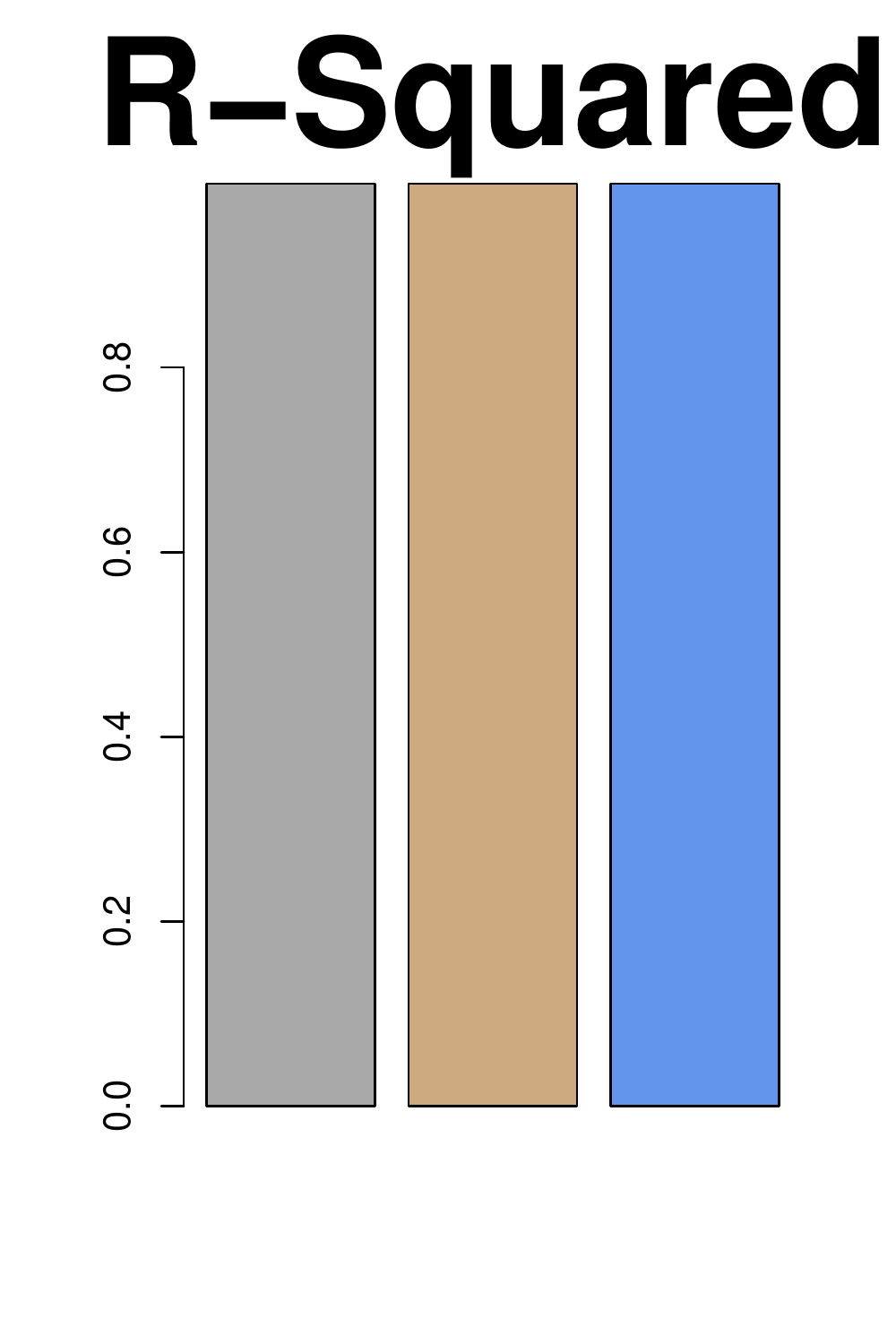} &
		\includegraphics[scale=0.3]{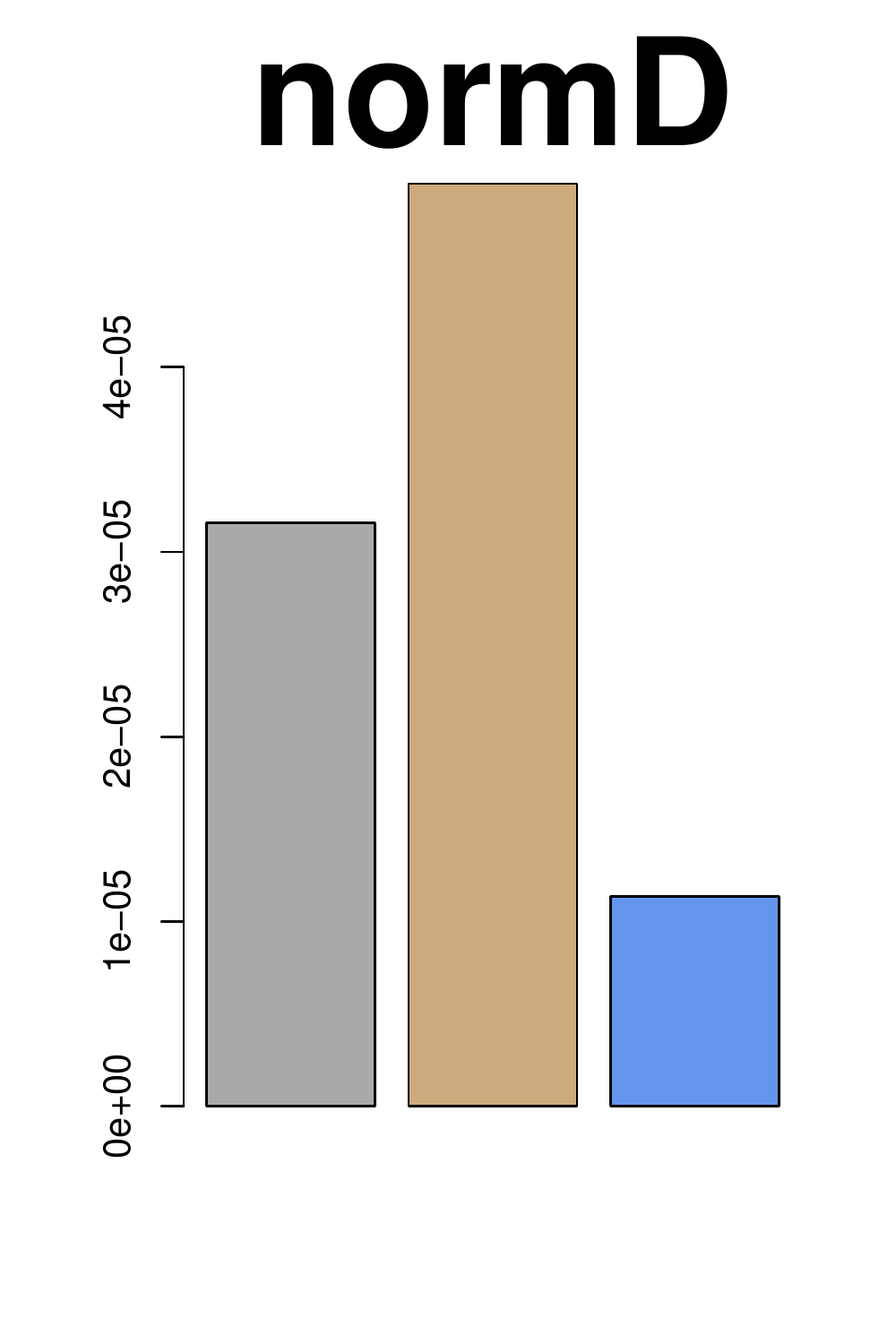} 
	\end{tabular}
	\caption{Easom Function: Performance comparison between modified HM (gray), MSCE with $N = 50$ (brown), and MSCE with $N = 230$ (blue).} 
	\label{Fig: easom1}
\end{figure}

From Table~\ref{Tab: easom1} and Figure~\ref{Fig: easom1}, we see that the proposed MSCE method using a budget size of $N = 230$ is clearly the most favoured approach. As compared to the modified HM approach (Bhattacharjee et al. 2019), the proposed method shows an improvement margin of $(2.46 - 1.48)/1.48 \times 100\% \approx 66\%$ per RMSE, and $(3.157- 1.137)/1.137 \times 100\% \approx 178\%$ according to log-normalized discrepancy.

\subsection{Example 2: Harari and Steinberg, 2014}
{\rm 
	In another example of applying the proposed MSCE approach to solve the inverse problem, we use a more complex test function by Harari and Steinberg (2014) with the  three-dimensional input space scaled to unit cube, i.e., $x = (x_1, x_2, x_3)^T \in [0,1]^3$.  The simulator model output is generated as:
	\begin{eqnarray*}
		y_t(x)= \exp(3x_1t + t)\times \cos(6x_2t+ 2t - 8x_3 - 6),
	\end{eqnarray*}
	where time $t \in [0,1]$ is on a 200-point equidistant grid. The target time-series response for the calibration problem corresponds to $x_0 = (0.522, 0.950, 0.427)^T$ (as in Example~4 of Zhang et al., 2019).

	Using the spline based knot selection method presented in Section~3,  we identified the DPS to be $\{118, 26, 95\}$. An initial training set of size $n_0 = 18$ generated via a MaxPro Latin Hypercube design and a total budget of $N = 66$  was needed when using a predetermined cutoff of $c = 3$. The budget used in implementing the modified history matching method is $N = 93$, thus the proposed method with the matching budget and $n_0$ held constant was used as well.  Figure~\ref{Fig:hs_a} depicts the placement of DPS and the time series responses for the inverse solutions derived from the methods.

	\begin{figure}[h!]
		\begin{center}
			\begin{tabular}{cc}
				\includegraphics[scale=.6]{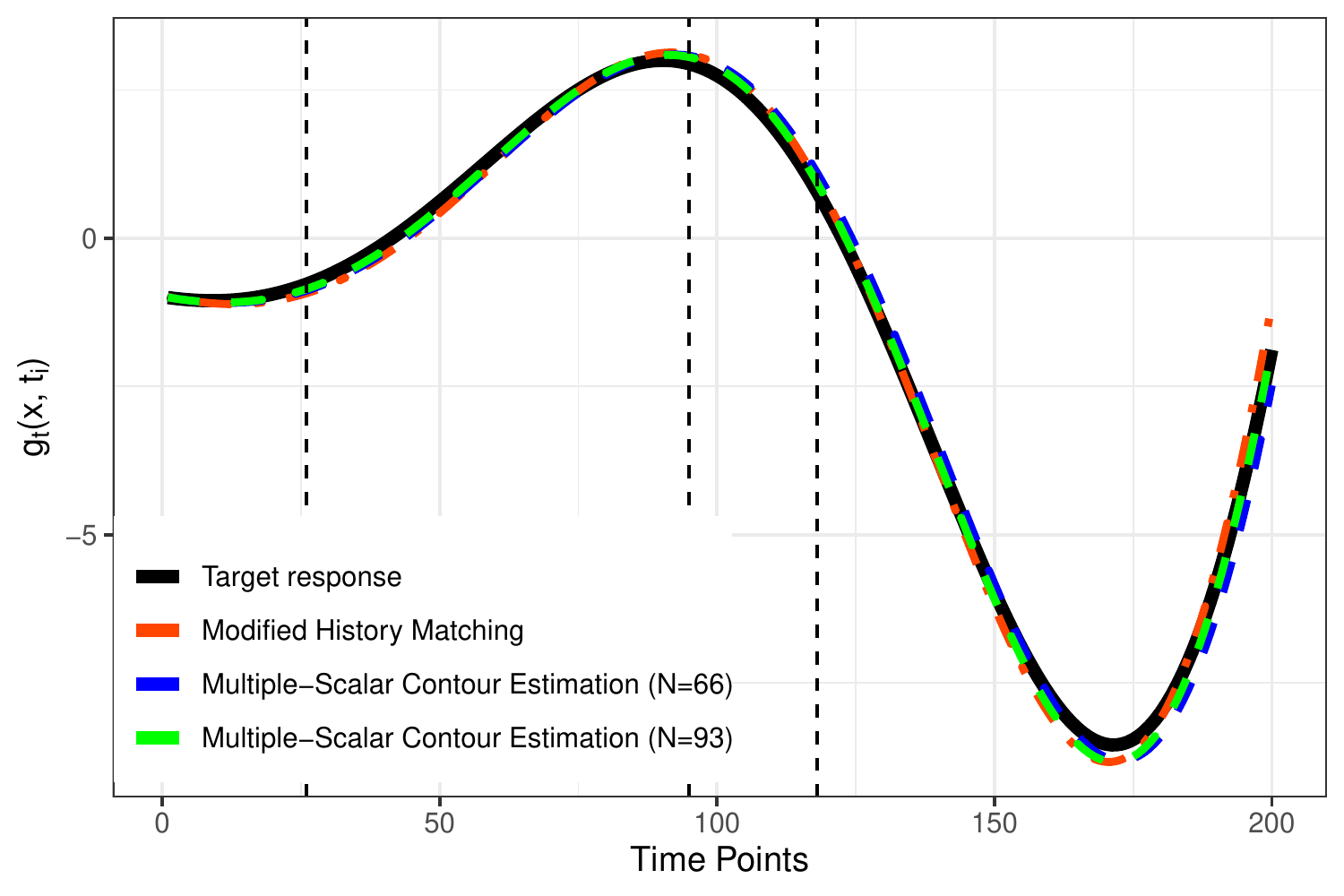} 
			\end{tabular}
			\caption{Harari and Steinberg (2014): The target series corresponding to $x_0 = (0.522, 0.950, 0.427)^T$ is shown by the black curve, and DPS is displayed by the dotted vertical lines. The time series responses corresponding to the estimated solution via modified HM method, MSCE method with $N=66$, and MSCE method with $N=93$ are shown by the red, blue, and green lines, respectively.}
			\label{Fig:hs_a}
		\end{center}
	\end{figure}
	
	Table~\ref{Tab: hs1} and  Figure~\ref{Fig: hs1} present the performance comparison of the two methods for this example.
	
	\begin{table}[h!]
		\centering
		\caption{Harari and Steinberg (2014): Performance comparisons of the proposed  MSCE using $N=66$ and $N=93$, and the modified HM method - which required 93 simulator runs.} 
		\medskip
		\begin{tabular}{llccc}
			\hline
			\bf{Methods} & \bf{Total Budget} & \bf{RMSE} & \bf{R$^2$} &  \bf{normD}\\
			\hline
			HM & $N = 93$ & 0.209 & 0.997 &  0.00317\\
			MSCE & $N = 66$ & 0.246 & 0.996 & 0.00440\\
			MSCE & $N = 93$ & 0.134 & 0.999 & 0.00132\\
			\hline
		\end{tabular}
		\label{Tab: hs1}
	\end{table}

	\begin{figure}[h!]
		\centering
		\begin{tabular}{lll}
			\includegraphics[scale=0.3]{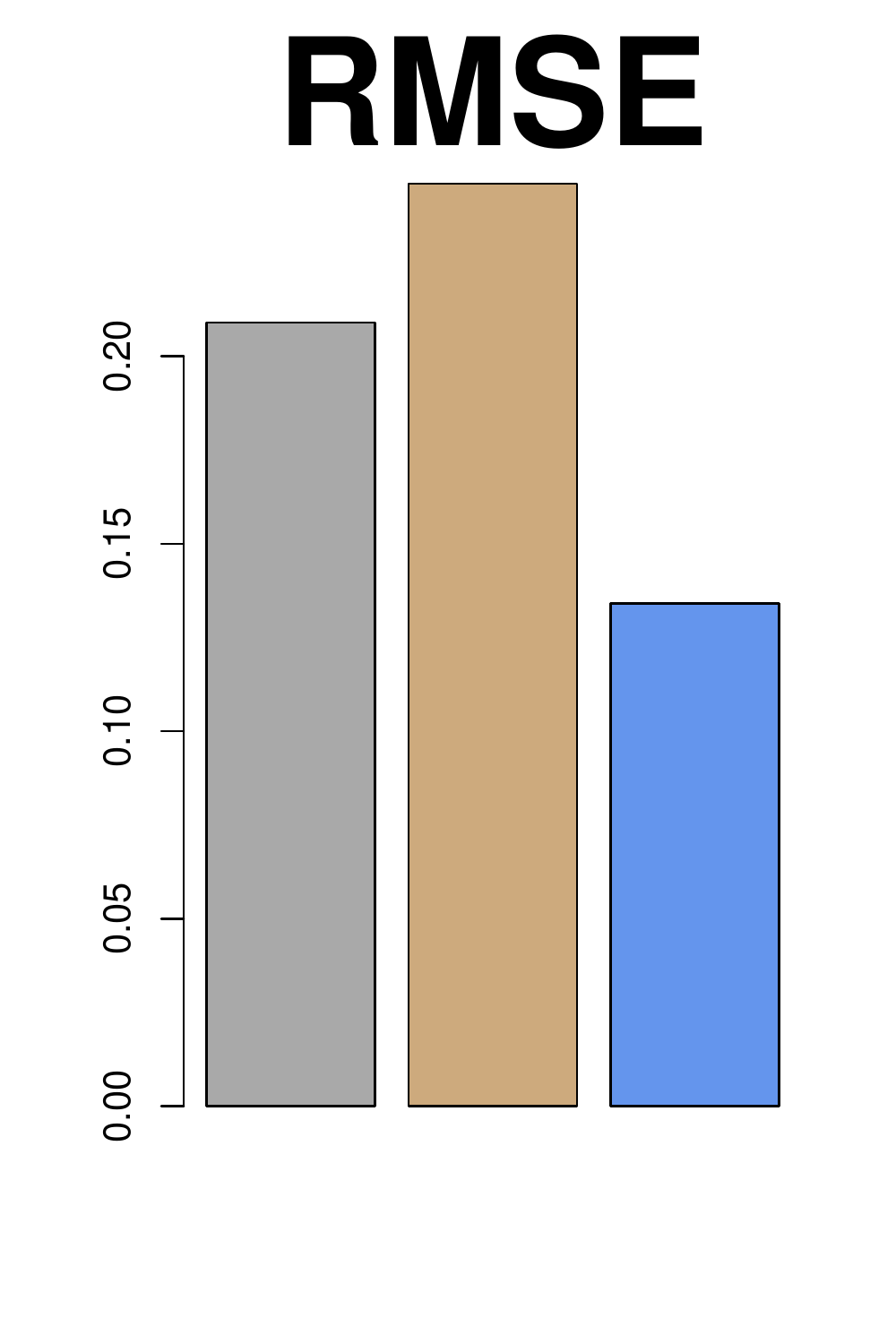} &
			\includegraphics[scale=0.3]{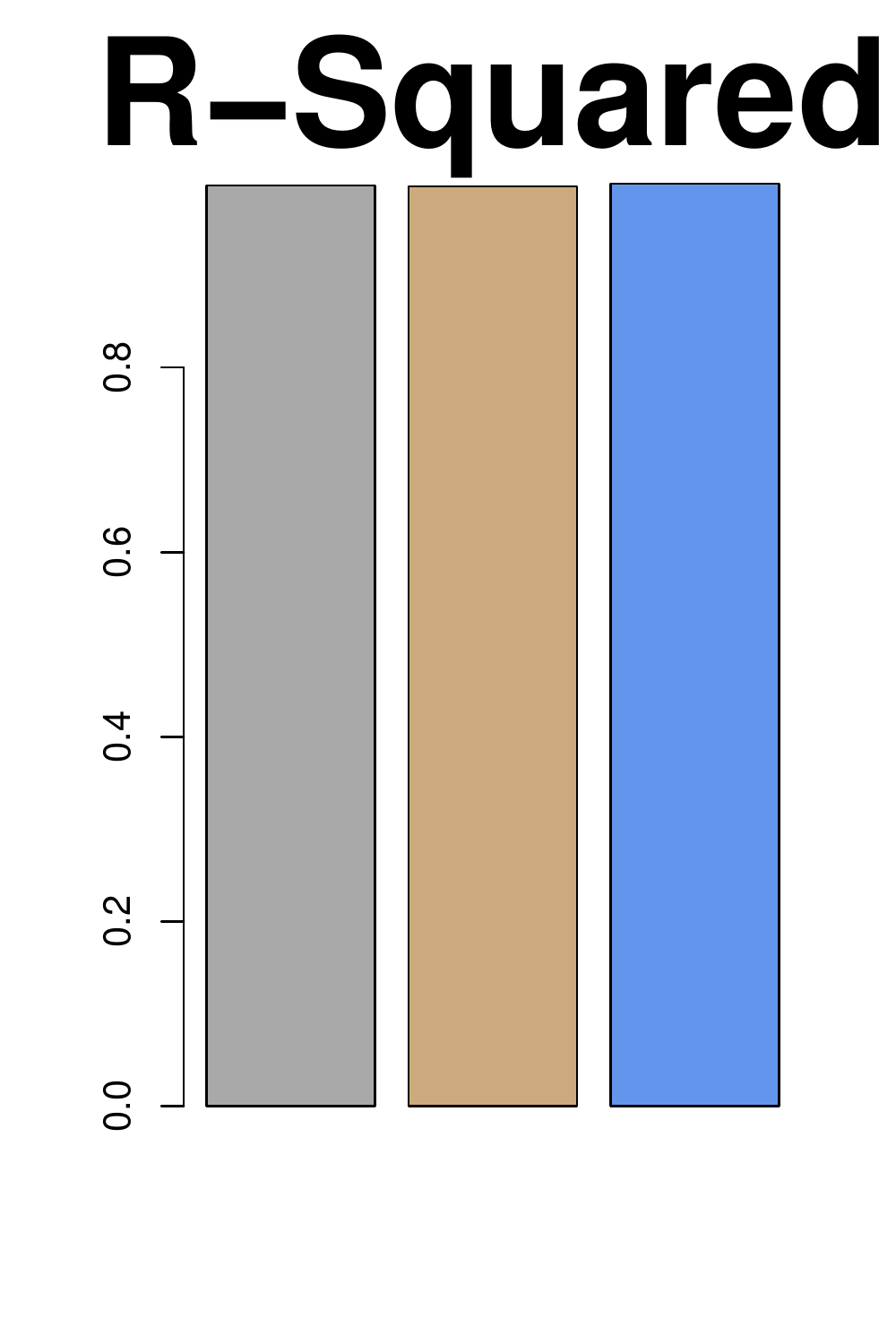} &
			\includegraphics[scale=0.3]{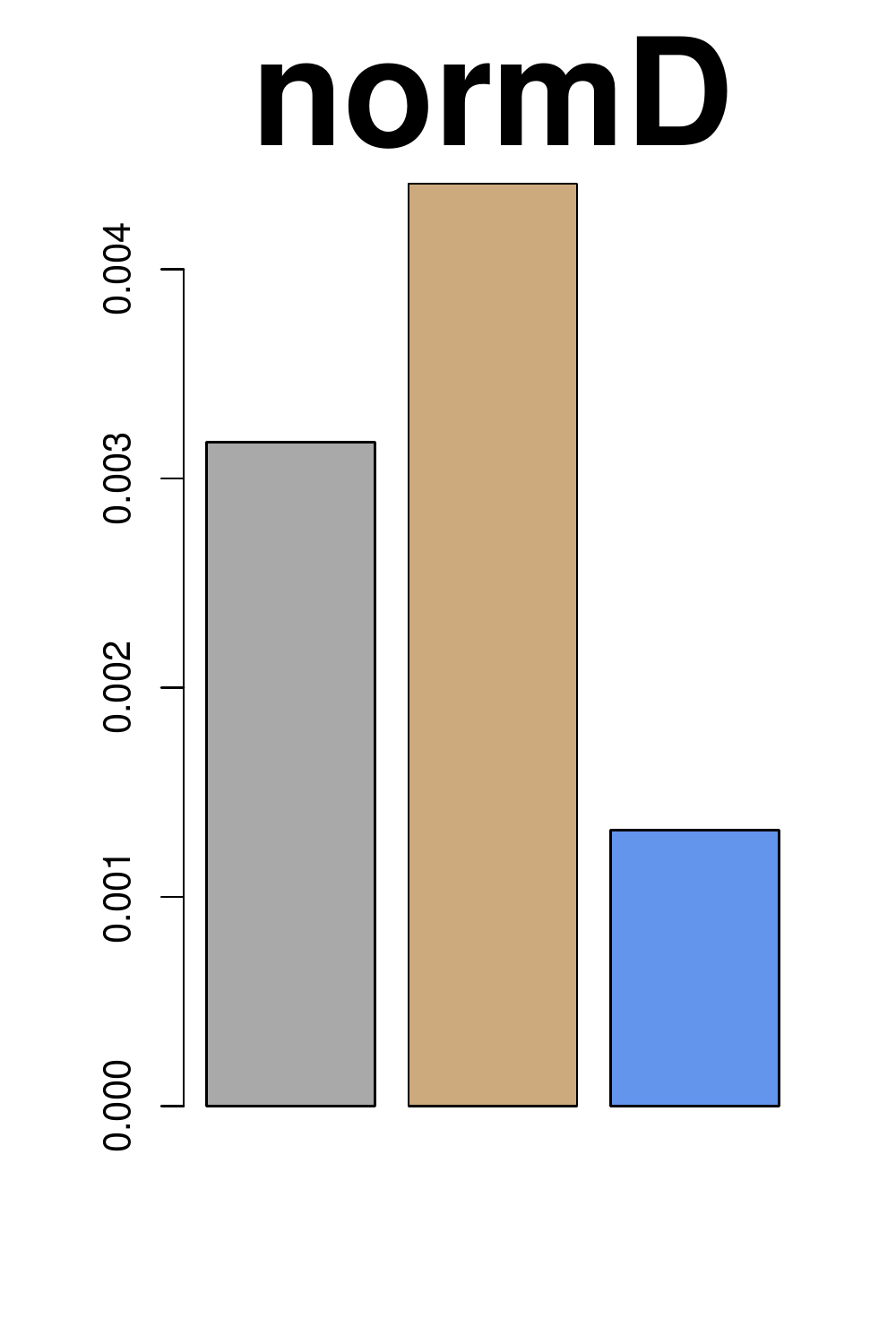} 
		\end{tabular}
		\vspace{-1cm}
		\caption{Harari and Steinberg (2014): Performance comparison between modified HM (with $N=93$ points) (in gray), MSCE with $N = 66$ (brown), and MSCE with $N = 93$ (blue).} 
		\label{Fig: hs1}
	\end{figure}

	From Table~\ref{Tab: hs1} and Figure~\ref{Fig: hs1} we see that the proposed MSCE method at the default budget of $N=66$ performs slightly worse than the modified HM method, however, the proposed method outperforms the modified HM method in all three goodness of fit measurements when the budget matches. In particular, we see the proposed method using $N=93$ outperform the modified HM method by a significant margin of $(0.209 - 0.134)/0.134 \times 100\% \approx 56\%$ according to RMSE, and an even greater margin of $(0.00317 - 0.00132)/0.00132 \times 100\% \approx 140\%$ according to log-normalized discrepancy.
}

\subsection{Example 3: Bliznyuk et al., 2008}
{\rm 
	We now use a five-dimensional environmental model by Bliznyuk et al. (2008) which simulates a pollutant spill caused by a chemical accident. Here, the input space is $x = (x_1, x_2, x_3, x_4, x_5)^T \in [7,13] \times [0.02, 0.12] \times [0.01, 3] \times [30.01, 30.304] \times [0,3]$, and the simulator outputs are generated as:
	\begin{eqnarray*}
		y_t(x) &=& \frac{x_1}{\sqrt{x_2t}} \exp\bigg(\frac{-x_5^2}{4x_2t}\bigg)  +  \frac{x_1}{\sqrt{x_2(t - x_4)}}\exp\bigg(\frac{-(x_5-x_3)^2}{4x_2(t - x_4)}\bigg)I(x_4<t)  
	\end{eqnarray*}
	where $t \in [35.3, 95]$ is defined by 200 equidistant points. The true target time-series response corresponds to $x_0 = (9.640, 0.059, 1.445, 30.277, 2.520)^T$ (as in Example~5 of Zhang et al., 2019). For the purpose of HM procedure, the input space is scaled such that $x \in [0,1]^5$.
	
	By optimizing the cubic spline knots onto the target response, the DPS of size $k=3$ is chosen at $\{30, 7, 61, 14\}$. An initial training set of size $n_0 = 30$ and total budget of $N=120$ are used for implementing the proposed MSCE method. Since the HM method (with $c = 0.7$) required $N=269$ simulator runs, the proposed method has also been implemented using $N=269$ for the calibration problem. Figure~\ref{Fig:bliz_a} shows the target response, the DPS, and the estimated inverse solution obtained by the three procedures.

	\begin{figure}[h!]
		\begin{center}
			\begin{tabular}{cc}
				\includegraphics[scale=0.6]{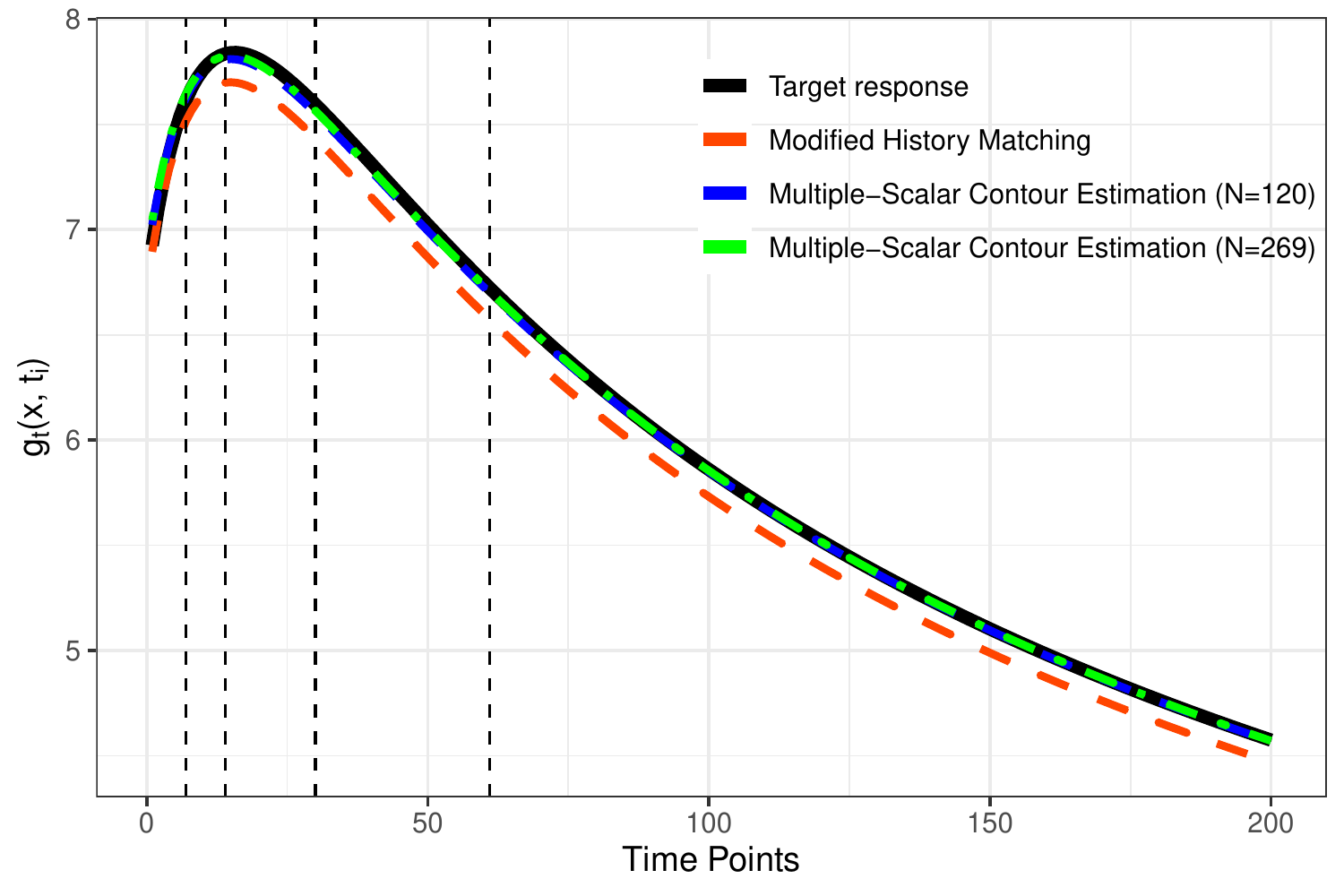} 
			\end{tabular}
			\caption{Bliznyuk et al. (2008): Target series is shown by the black curve, and DPS of size $k = 4$ is displayed by the dotted vertical lines. The estimated inverse solutions obtained via the modified HM method, MSCE method with $N=120$ runs, and MSCE method with $N=269$ runs are shown by the red, blue, and green lines respectively.}
			\label{Fig:bliz_a}
		\end{center}
	\end{figure}
	
}

\begin{figure}[h!]
	\centering
	\begin{tabular}{ccc}
		\includegraphics[scale=0.3]{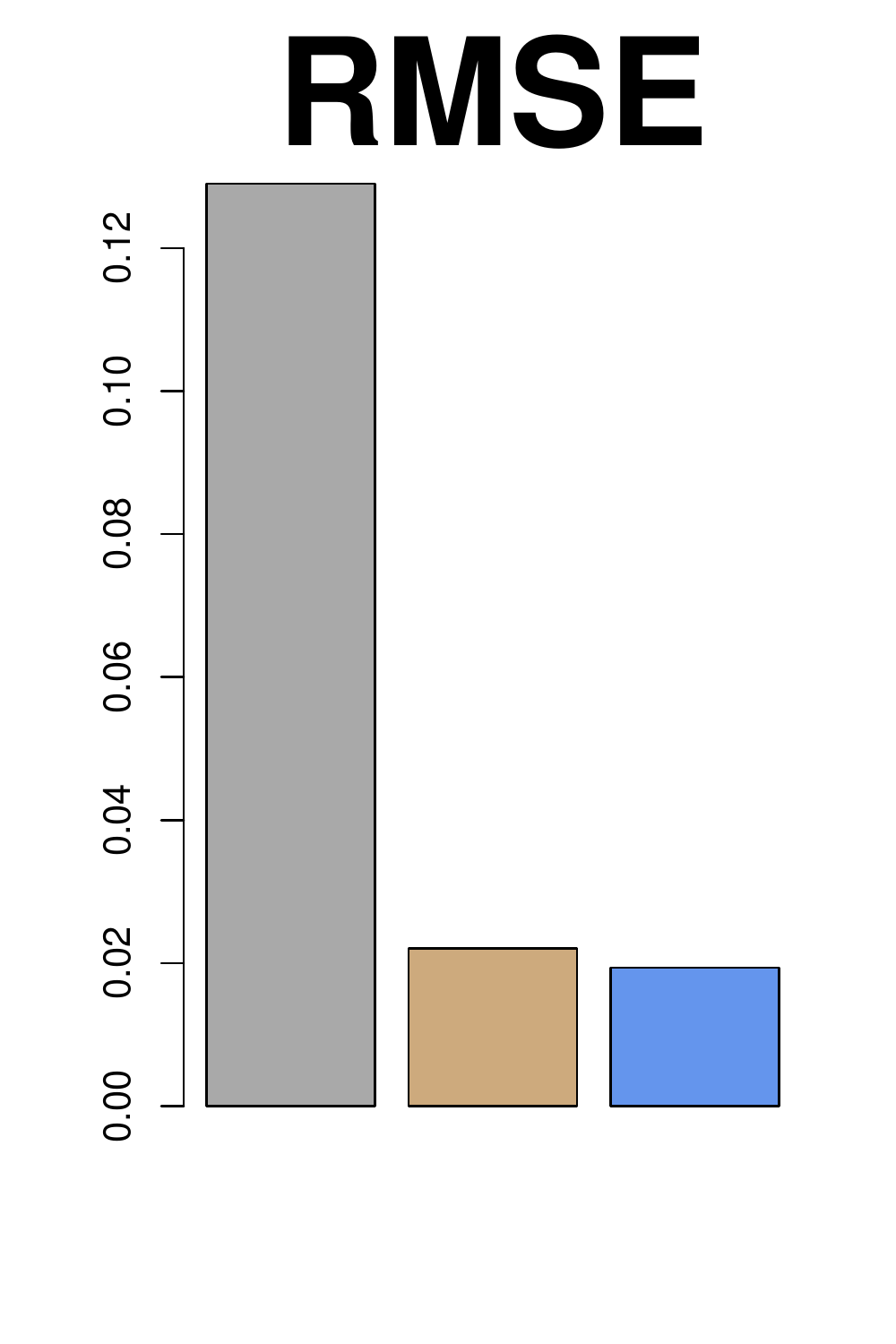} &
		\includegraphics[scale=0.3]{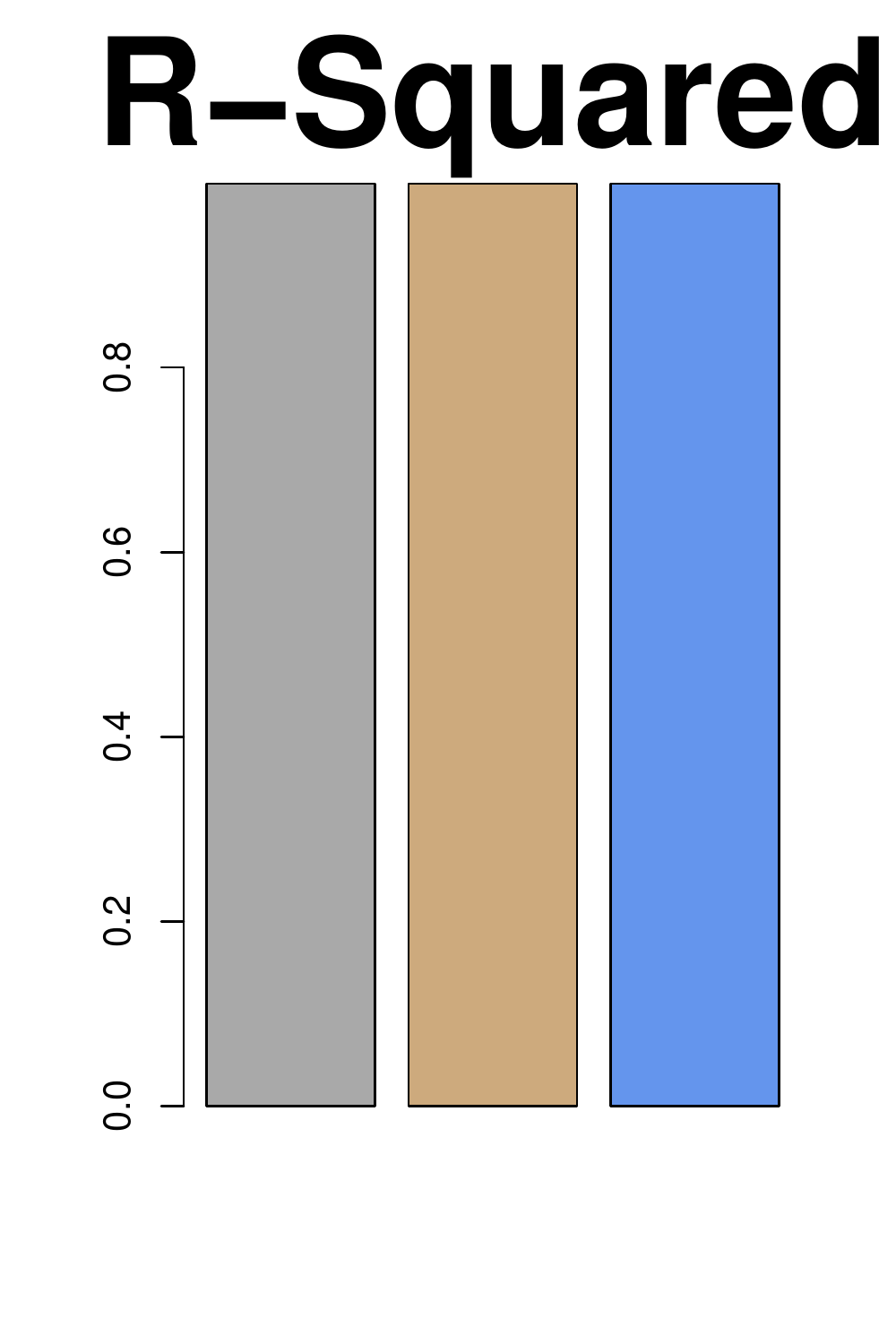} &
		\includegraphics[scale=0.3]{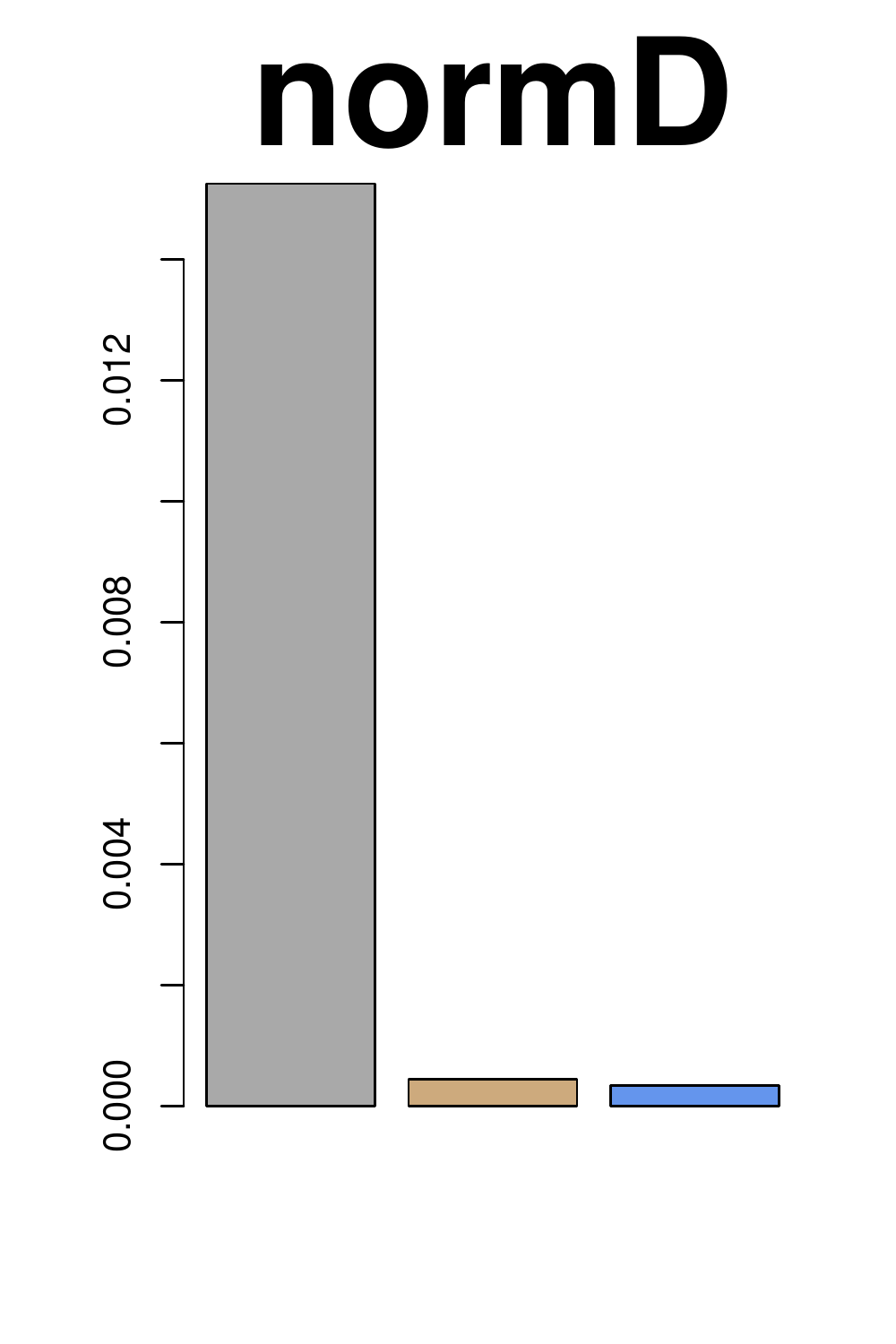} 
	\end{tabular}
	\vspace{-1cm}
	\caption{Bliznyuk et al. (2008): Performance comparison plots for modified HM algorithm with $N = 269$ (gray), MSCE with $N = 120$ (brown), and MSCE with $N = 269$ (blue).} 
	\label{Fig: bliz1}
\end{figure}

From Figure~\ref{Fig:bliz_a}, it is clearly visible that the HM algorithm leads to inferior inverse solution as compared to the proposed MSCE approaches. Table~\ref{Tab: bliz1} and  Figure~\ref{Fig: bliz1} compare the goodness of fit measures for this five-dimensional example.

\begin{table}[h!]
	\centering
	\caption{Bliznyuk et al. (2008): Performance comparison of the proposed MSCE method using $N=120$ and $N=269$, and the modified HM method with $N=269$ simulator runs.} 
	\medskip
	\begin{tabular}{llccccccc}
		\hline
		\bf{Methods} &&& \bf{Total Budget} && \bf{RMSE} & \bf{R$^2$} & \bf{normD}\\
		\hline
		modified history matching &&& $N = 269$ && 0.129 & 0.999 &  0.0152\\
		multiple scalar contour estimation &&& $N = 120$ && 0.0221 &  0.999 & 0.000447\\
		multiple scalar contour estimation &&& $N = 269$ &&  0.0194 & 0.999 &  0.000346\\
		\hline
		%		\% improvement (HM $\rightarrow$ EI) &&  &  &  &  & \\
		%		\hline
		%		\% budget increase (HM $\rightarrow$ EI) &  -124.17\% &  &  &  &  & \\
		%		\hline
	\end{tabular}
	\label{Tab: bliz1}
\end{table}

Table~\ref{Tab: bliz1} and  Figure~\ref{Fig: bliz1} show that the proposed multiple scalar contour estimation (MSCE) method with both budget levels ($N=120, N=269$) significantly outperform the modified history matching approach by massive margins.

\section{Real Application: Rainfall-Runoff Example}
The modified HM model proposed by Bhattacharjee et al. (2019) motivated their approach using hydrological models. The actual computer simulator - Matlab-Simulink model introduced by Duncan et al. (2013)  - studies the rainfall-runoff relationship for the windrow composting pad. The following four input parameters are identified to have the most significant influence on the output: depth of surface, depth of sub-surface and two coefficients of the saturated hydraulic conductivity ($K_{sat1}$ and $K_{sat2}$). Interested readers can see Duncan et al. (2013) for further details on the Matlab-Simulink Model.

\begin{figure}[h!]
	\begin{center}
		\begin{tabular}{cc}
			\includegraphics[scale=0.7]{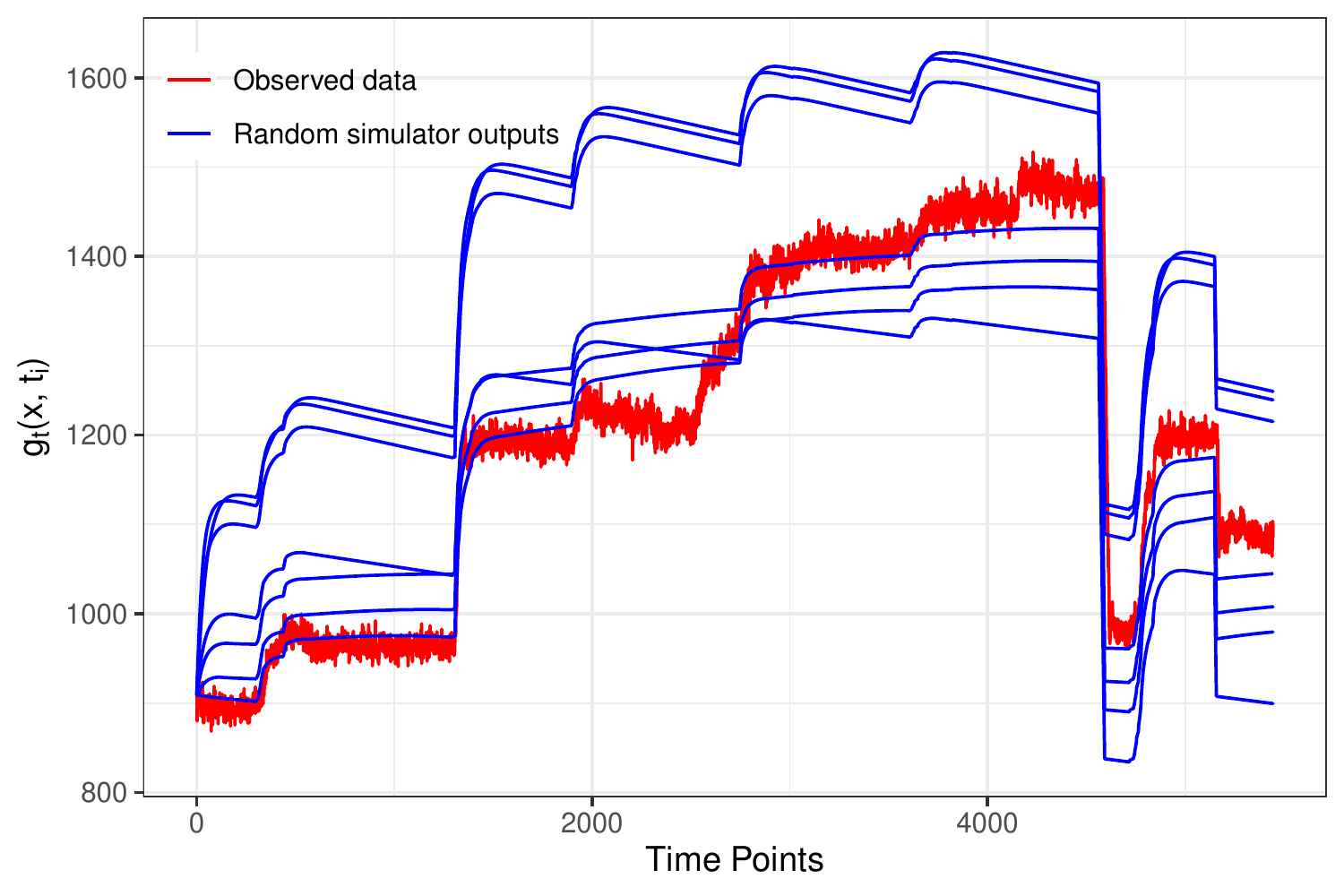} 
		\end{tabular}
		\caption{Matlab-Simulink model: Observed data $(g_0)$ with a few random simulator outputs.}
		\label{ms_random}
	\end{center}
\end{figure}

For the inverse problem, the target response is the rainfall-runoff data ($g_0$) observed from the Bioconversion center at the University of Georgia, Athens, USA (Bhattacharjee et al. 2019). Figure~\ref{ms_random} depicts the observed target response and a few random outputs from the Matlab-Simulink model.

\begin{figure}[h!]
	\begin{center}
		\begin{tabular}{cc}
			\includegraphics[scale=0.7]{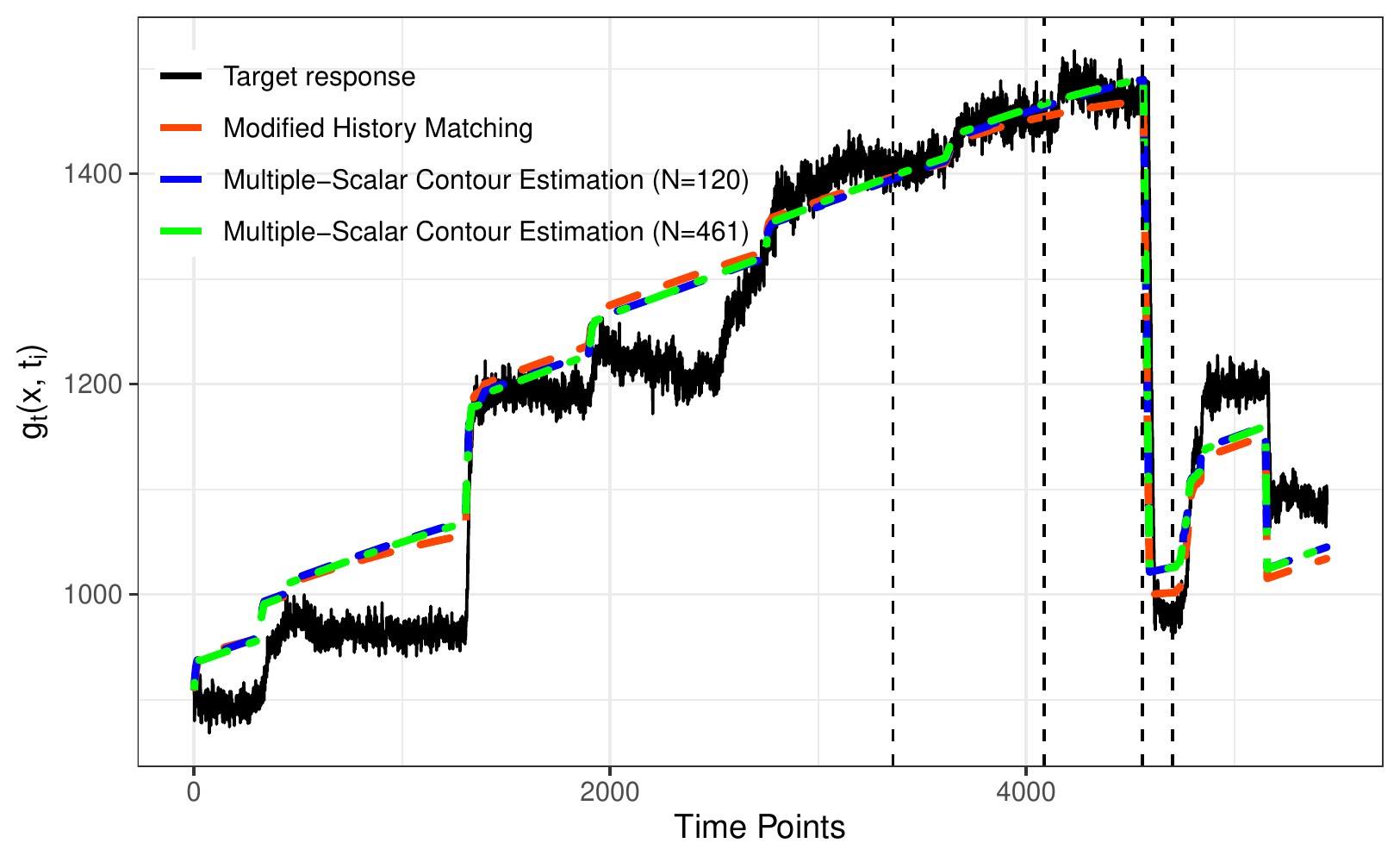} 
		\end{tabular}
		\caption{Matlab-Simulink model: Target time series response is shown by the black curve, and the DPS is being displayed by the dotted vertical lines. The estimated inverse solution corresponding to the modified HM method, MSCE method using $N=120$, and MSCE method using $N=461$ are displayed by the red, blue, and green lines respectively.}
		\label{Figs:ms_a}
	\end{center}
\end{figure}

Using the spline-based knots selection method discussed in Section~3,  we found the discretization time points as $DPS = \{4557, 3359, 4702, 4085\}$. For the proposed approach, we used an initial training set with $n_0 = 40$ points generated using a Latin hypercube design and a total budget of $N= 120$. However, since the modified HM method exhausted a total budget of $N = 461$ using a predetermined cutoff of $c = 3$ (as in Bhattercharjee et al., 2019), the budget size of $N = 461$ was also made available for the proposed method. Figure~\ref{Figs:ms_a} shows the positioning of DPS as well as the estimated time-series responses for the three inverse solutions with respect to the observed runoff data.

\begin{figure}[h!]
	\centering
	\begin{tabular}{ccc}
		\includegraphics[scale=0.3]{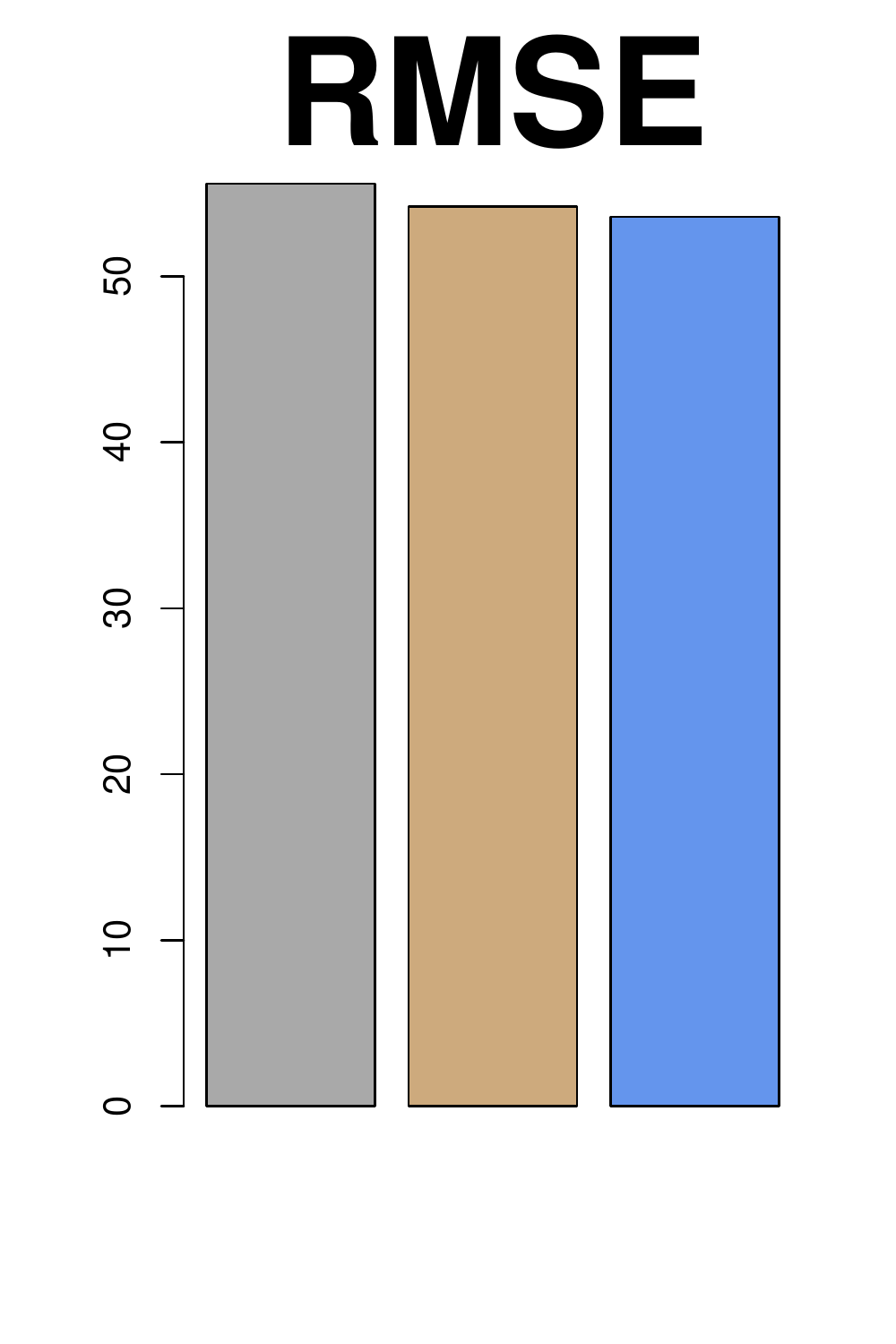} &
		\includegraphics[scale=0.3]{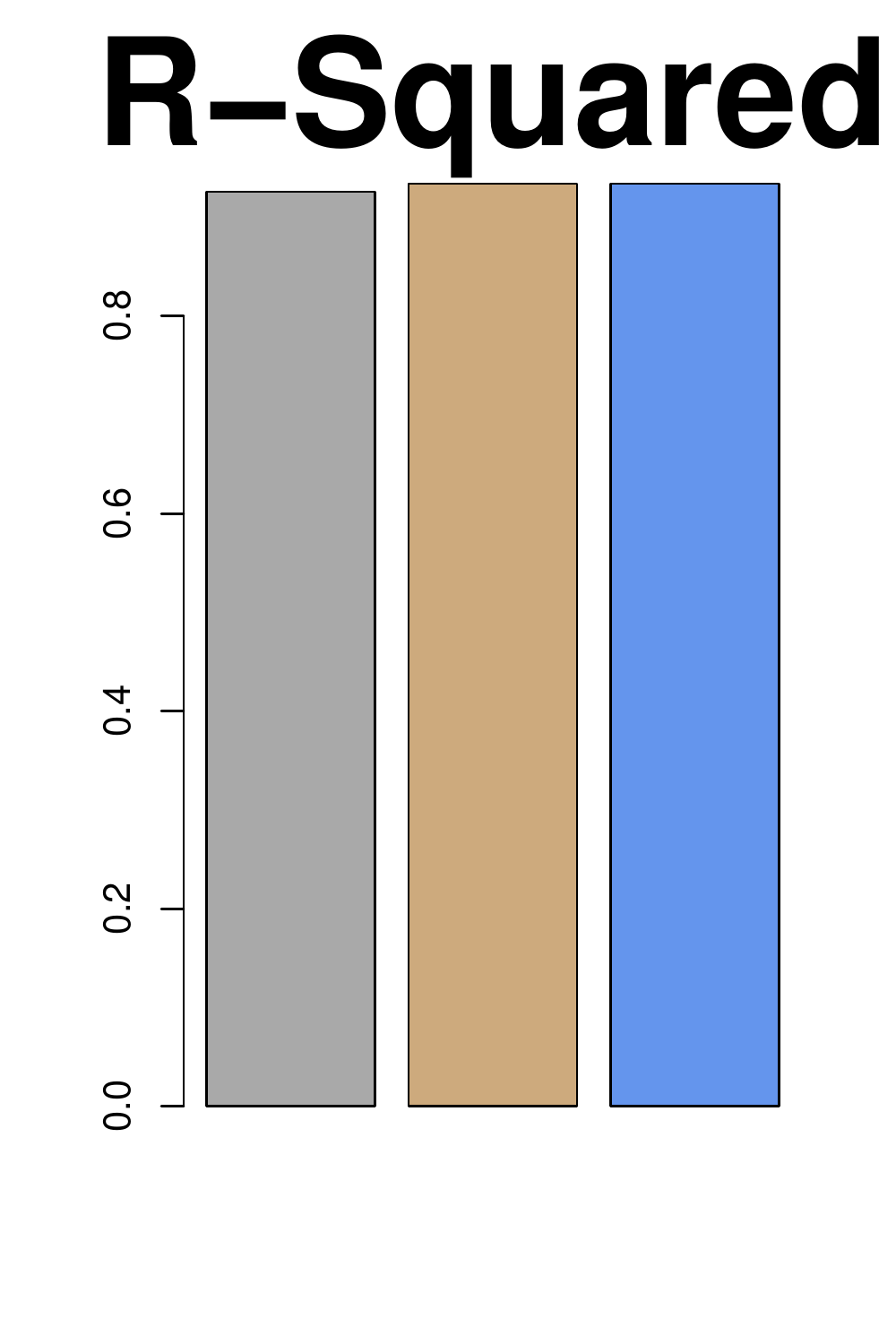} &
		\includegraphics[scale=0.3]{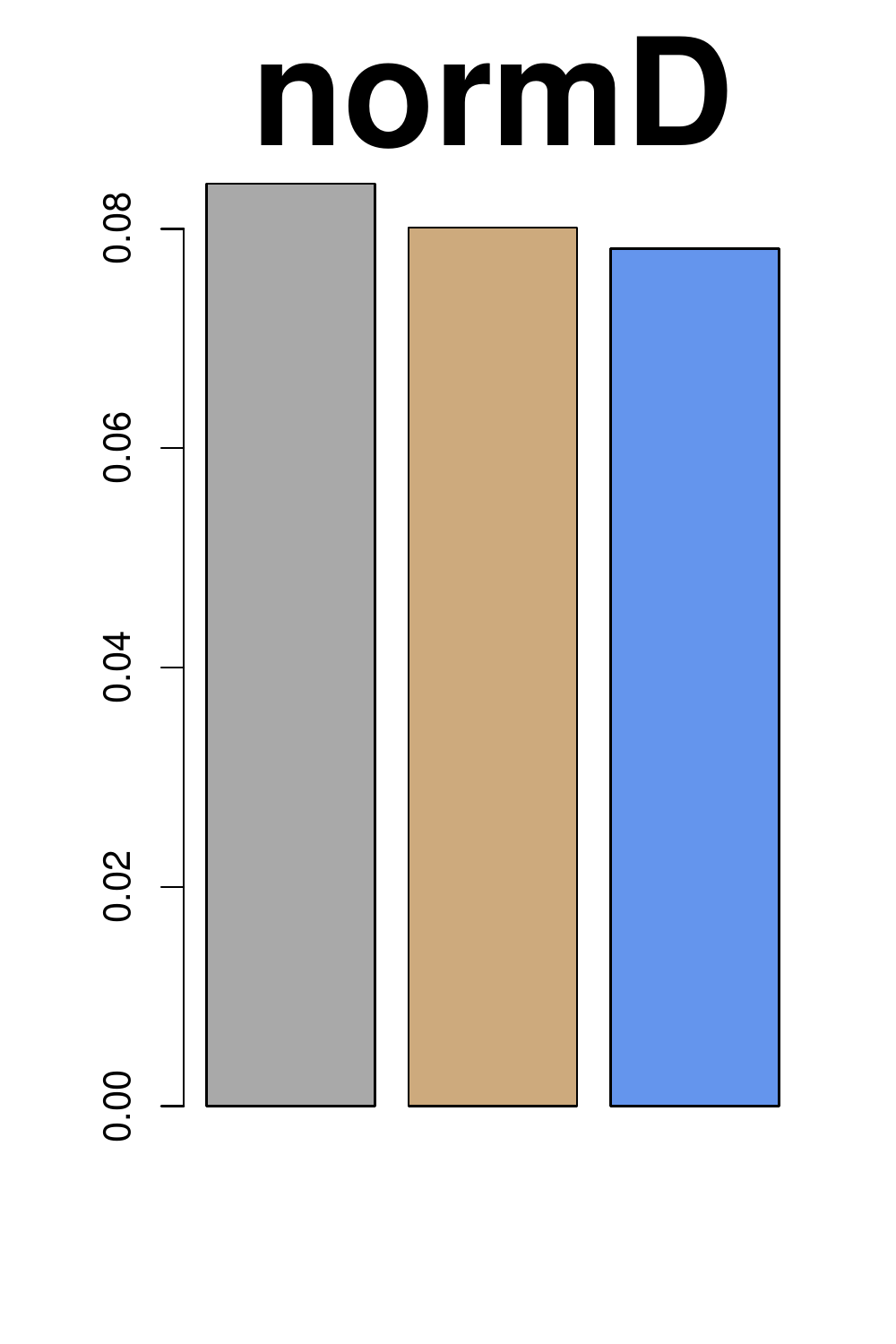} 
	\end{tabular}
	\vspace{-1cm}
	\caption{Matlab-Simulink model: Pictorial representation of performance comparison between modified HM with $N = 461$ (gray), MSCE with $N = 120$ (brown), and MSCE with $N = 461$ (blue).} 
	\label{Fig: ms1}
\end{figure}

Table~\ref{Tab: ms1} and Figure~\ref{Fig: ms1} show that the MSCE approach using $N=120$ outperformed the modified HM approach by small margins of $(55.580 - 54.215)/54.215 \times 100\% \approx 2.5\%$ per RMSE, and $(0.0841 - 0.0801)/0.0801 \times 100\% \approx 5\%$ per log-normalized discrepancy. When the proposed method matches the budget size used by modified HM method at $N = 461$, the margin of improvement increases to $(55.580 - 53.585)/53.585 \times 100\% \approx 3.7\%$ per RMSE, and $(0.0841 - 0.0782)/0.0782 \times 100\% \approx 7.5\%$ per normal discrepancy.

\begin{table}[h!]
	\centering
	\caption{Matlab-Simulink model: Goodness-of-fit comparisons of the proposed MSCE method using $N=120$ and $N=461$, and the modified HM method (with $N=461$ runs).} 
	\medskip
	\begin{tabular}{cccccc}
		\hline
		\bf{Methods} & \bf{Total Budget} & \bf{RMSE} & \bf{R$^2$} & \bf{normD}\\
		\hline
		HM & $N = 461$ & 55.580 & 0.926  & 0.0841\\
		MSCE & $N = 120$ & 54.215 & 0.934  & 0.0801\\
		MSCE & $N = 461$ &  53.585 & 0.934  &  0.0782\\
		\hline
	\end{tabular}
	\label{Tab: ms1}
\end{table}

\section{Concluding Remarks}
In this paper, we have proposed a scalarized approach of solving the inverse problem for dynamic computer simulators by first carefully selecting a handful of time-points based on the target response series (called the DPS), and then solve multiple scalar-valued inverse problems at the DPS using the popular sequential algorithm via expected  improvement approach developed by Ranjan et al. (2008). The final inverse solution for the underlying dynamic simulator will be the intersection of all scalarized inverse solutions. In this paper, we have also proposed using a natural cubic spline based method for systematically finding the DPS (discretization point set). Based on the our investigation using several test functions and a real-life hydrological simulator, it is clear that the proposed approach outperforms the competing modified history matching algorithm by some or significant margin.  It has also been demonstrated that the proposed MSCE approach shows better performance than the modified HM algorithm even with a much lower simulator run budget. This is a two-fold advantage: in the accuracy of inverse solution and the number of simulator runs used. 

There are a few important remarks worth mentioning. First, when finding an optimal DPS using spline-based technique, we followed a greedy ``forward variable selection" type approach and identified one best knot at-a-time. Perhaps the ``best selection" type approach might lead to a better solution, however, it was abandoned due to vast (seemingly impractical) computational cost in finding the best DPS. Second, when solving the scalar-valued inverse problems at the $i$-th element of the DPS, it is important to note that the size of the initial design is $n_0 + (i-1)\cdot (N-n_0)/k$ and budget of follow-up points is $(N-n_0)/k$. Thus, it is tempting to believe that the inverse solution at the first or second element of DPS are perhaps less accurate as compared to (say) the fourth or fifth time point in the DPS. Based on our preliminary simulation study, we found no significant improvement in accuracy by changing the order of DPS for solving the scalarized inverse problems. On a related note, the distribution of total budget $N$ for solving $k$ scalar-valued inverse problems can also be further investigated. As per our preliminary simulation study, the results were not significantly different.

	%	\clearpage
	%	\section*{Acknowledgment}

	%%%%%%%%%%%%%%%%%%%%%%%%%%%%%%%%%%%%
	%                                                                                             %
	%                                                                                             %
	%                                                                                             %
	%                                                                                             %
	%                                                                                             %
	%                             SECTION 6: References                            %
	%                                                                                             %
	%                                                                                             %
	%                                                                                             %
	%                                                                                             %
	%                                                                                             %
	%%%%%%%%%%%%%%%%%%%%%%%%%%%%%%%%%%%%
	%\clearpage
	\section*{References}\label{Ref}
	\begin{enumerate}
		\item
		Bhattacharjee, N. V., Ranjan, P., Mandal, A., \& Tollner, E. W. (2019). A history matching approach for calibrating hydrological models. \emph{Environmental and Ecological Statistics, 26}(1), 87-105.
		
\item
		Bingham, D., Ranjan, P., \& Welch, W. J. (2014). Design of computer experiments for optimization, estimation of function contours, and related objectives. \emph{Statistics in Action: A Canadian Outlook, 109.}
		
\item
		Bliznyuk, N., Ruppert, D., Shoemaker, C., Regis, R., Wild, S., \& Mugunthan, P. (2008). Bayesian calibration and uncertainty analysis for computationally expensive models using optimization and radial basis function approximation. \emph{Journal of Computational and Graphical Statistics, 17}(2), 270-294.
		
\item
		Duncan O, Tollner E, Ssegane H (2013) An instantaneous unit hydrograph for estimating runoff from windrow composting pads. \emph{Appl Eng Agric 29}(2):209–223
		
\item
		Franke, R. (1979). A critical comparison of some methods for interpolation of scattered data. \emph{NAVAL POSTGRADUATE SCHOOL MONTEREY CA NPS53-79-003.}
		
\item
		Harari, O., \& Steinberg, D. M. (2014). Convex combination of Gaussian processes for Bayesian analysis of deterministic computer experiments. \emph{Technometrics, 56}(4), 443-454.
		
\item
		Johnson, M. E., Moore, L. M., \& Ylvisaker, D. (1990). Minimax and maximin distance designs. \emph{Journal of statistical planning and inference, 26}(2), 131-148.
		
\item
		Jones, D. R., M. Schonlau, and W. J. Welch (1998). Efficient global optimization of expensive black-box functions. \emph{Journal of Global Optimization 13}(4), 455–492.
		
\item
		Joseph, V. R., Gul, E., \& Ba, S. (2015). Maximum projection designs for computer experiments. \emph{Biometrika, 102}(2), 371-380.
		
\item
		MacDonald, B., Ranjan, P., \& Chipman, H. (2015). GPfit: An R package for fitting a Gaussian process model to deterministic simulator outputs. \emph{Journal of Statistical Software, 64}(i12).
		
\item
		Michalewicz, Z. (1996), \emph{Genetic Algorithms+Data Structures $\frac{1}{4}$ Evolution Programs}, SpringerVerlag, Berlin/Heidelberg/New York.
		
\item
		Nash, J. E., \& Sutcliffe, J. V. (1970). River flow forecasting through conceptual models part I—A discussion of principles. \emph{Journal of hydrology, 10}(3), 282-290.
		
\item
		Picheny, V., Ginsbourger, D., Richet, Y., \& Caplin, G. (2013). Quantile-based optimization of noisy computer experiments with tunable precision. \emph{Technometrics, 55}(1), 2-13.

\item
		Ranjan, P., Bingham, D., \& Michailidis, G. (2008). Sequential experiment design for contour estimation from complex computer codes. \emph{Technometrics, 50}(4), 527-541.
		
\item
		Ranjan, P., Thomas, M., Teismann, H., \& Mukhoti, S. (2016). Inverse problem for a time-series valued computer simulator via scalarization. \emph{Open Journal of Statistics, 6}(3), 528-544.
		
\item
		Sacks, J., Welch, W. J., Mitchell, T. J., \& Wynn, H. P. (1989). Design and analysis of computer experiments. \emph{Statistical science}, 409-423.
		
\item
		Vernon, I., Goldstein, M., \& Bower, R. G. (2010). Galaxy formation: a Bayesian uncertainty analysis. \emph{Bayesian analysis, 5}(4), 619-669.
		
\item
		Zhang, R., Lin, C. D., \& Ranjan, P. (2018). A sequential design approach for calibrating a dynamic population growth model. \emph{arXiv preprint arXiv:1811.00153.}
	\end{enumerate}

\end{document}